\documentclass[aps,pra,twocolumn,amsmath,amssymb,floatfix,longbibliography]{revtex4-2} %

\usepackage{graphicx}
\usepackage{bm}
\usepackage{color}
\usepackage{color}
\usepackage[colorlinks, linkcolor=mycol, citecolor=mycol, urlcolor=mycol, breaklinks]{hyperref}

\definecolor{mycol}{RGB}{10,55,130}

\newcommand {\at} {\mathrm{at}}

\newcommand {\diff} {\mathrm{diff}}

\begin{document}

\title{Subradiance in dilute atomic ensembles: Role of pairs and multiple scattering}

\author{Y. A. Fofanov${}^{1}$  and I. M. Sokolov${}^{1,2}$\\
{\small $^{1}$Institute for Analytical Instrumentation, Russian Academy of
Sciences, 198095 St. Petersburg, Russia }\\
{\small $^{2}$ Peter the Great St. Petersburg Polytechnic University, 195251 St. Petersburg, Russia}\\
R. Kaiser and W. Guerin\\
{\small Universit\'{e} C\^{o}te d'Azur, CNRS, Institut de Physique de Nice, 06560 Valbonne, France}}

\date{\today}

\sloppy

\begin{abstract}
We study numerically the slow (subradiant) decay of the fluorescence of motionless atoms after a weak pulsed excitation. We show that, in the linear-optics regime and for an excitation detuned by several natural linewidths, the slow decay rate can be dominated by close pairs of atoms (dimers) forming superradiant and subradiant states. However, for a large-enough resonant optical depth and at later time, the dynamics is dominated by collective many-body effects. In this regime, we study the polarization and the spectrum of the emitted light, as well as the spatial distribution of excitation inside the sample, as a function of time during the decay dynamics. The behavior of these observables is consistent with what would be expected for radiation trapping of nearly resonant light.  This finding sheds light on subradiance in dilute samples by providing an interpretation based on the light behavior of the system (multiple scattering) which is complementary to the more commonly used picture of the collective atomic Dicke state. 
\end{abstract}

\maketitle

\section{Introduction}

Collective effects in light-atom interactions are at the focus of intense research, in particular for the potential applications to quantum optics and photonics \cite{Chang:2018}. In this context, subradiance, corresponding to the reduced spontaneous emission of some collective modes \cite{Dicke:1954, Pavolini:1985, Bienaime:2012}, has received considerable attention, with several experimental demonstrations \cite{Guerin:2016a, Jenkins2017, Solano:2017, Rui:2020, Ferioli:2021}. The long subradiant lifetimes may be used for storage and retrieval of quantum information \cite{Scully:2015, Facchinetti:2016, Jen:2016, Asenjo:2017, Ferioli:2021} and other photonic devices, for instance, single-atomic-layer mirrors \cite{Rui:2020} and nanolasers \cite{Holzinger:2020}.

Often, the term ``subradiance'' is used in a narrow sense as being due to only the near-field dipole-dipole interaction between atoms, like in the two-atom case ($N=2$) \cite{Stephen:1964, DeVoe:1996}. However, long-lived modes, as well as short-lived ones, also exist in macroscopic samples at low density, when the far-field part dominates the interaction. Since the short-lived modes are called superradiant \cite{Arecchi:1970, Rehler:1971, MacGillivray:1976, Svidzinsky:2008, Scully:2009, Araujo:2016, Roof:2016, Ortiz:2018}, it is consistent to call the long-lived ones subradiant. This collective subradiance ($N\gg 1$) in a broad sense has been predicted and observed recently in dilute samples \cite{Bienaime:2012, Guerin:2016a}. These terms are natural in the framework of a coupled-dipole picture, in which the light degrees of freedom are traced out and atoms interact with each other through the dipole-dipole interaction. Then the system properties can be described with eigenmodes and complex eigenvalues \cite{Svidzinsky:2008b, Svidzinsky:2010}, which allows one to classify the collective atomic states into two categories: superradiant and subradiant modes having decay rates larger and smaller than the single-atom one, respectively. Subradiance is thus a generic term for all long-lived modes.

However, it may also be desirable, for a better understanding, to develop other interpretations of super- and subradiance, which would rely on ``optical pictures,'' in which one would describe the electromagnetic field interacting with the atoms modeled as point-dipole scatterers and/or macroscopic effective quantities such as the refractive index. We recently provided such a description for superradiance in dilute samples in the linear-optics regime \cite{Weiss:2021}. Here we address the nature of the subradiant modes. Long lifetimes of light in an atomic sample can be due to different mechanisms, e.g., trapping due to refractive index boundaries or gradient \cite{Schilder:2016, Cottier:2018} and radiation trapping due to incoherent multiple scattering \cite{Labeyrie:2003}, with eventually mesoscopic corrections such as recurrent scattering, weak and strong localization effects \cite{Rossum:1999}, etc. Therefore, the problem is complicated in general.

In this article we restrict our study to the following case: The system is driven with a weak-intensity laser (linear-optics regime) and with a plane wave; the sample is three dimensional, statistically homogeneous, and dilute (the average interparticle distance is larger than the inverse of the wave number); and atoms are motionless and located at random positions. Moreover, the slow decay is studied after averaging over many realizations of the positional disorder.
We show that, in this regime of parameters, there are two main contributions to the slow decay at late time: recurrent scattering within diatomic clusters and radiation trapping. This is inferred from the computation, using the coupled-dipole model, of the temporal dynamics of the fluorescence, as done in several earlier works, but also using observables not studied in this context: the spatial distribution of atomic excitation and the spectrum and polarization of the fluorescence. Our findings provide a better understanding of the physics of subradiance in dilute disordered systems.

The paper is organized as follows. In the next section we present the coupled-dipole model and the parameters we use. In Sec. \ref{sec:results} we present our numerical results on the decay dynamics of the fluorescence, including spatial, spectral, and polarization properties. In Sec. \ref{sec:discuss} we summarize and discuss our findings and remaining open questions.

\section{Basic assumptions and approach}

\subsection{Coupled-dipole model}

We consider an ensemble consisting of $N \gg 1$ identical atoms with a nondegenerate ground state of angular momentum $J_\mathrm{g} = 0$. The excited  state is $J_\mathrm{e} = 1$. The lifetimes of all three of its Zeeman sublevels ($m=-1,0,1$) are the same and equal to $\tau_\at =1/\gamma$. We describe the evolution of the atomic states by the coupled-dipole (CD) model, which is traditional for this class of problems. This model was first proposed by Foldy \cite{Foldy:1945} and then discussed in detail by Lax \cite{Lax:1951}. Later a similar approach was used in the context of different types of collective effects such as multiple and recurrent scattering, collective spontaneous decay, and Anderson localization of light \cite{Javanainen:1999, Rusek:2000, Pinheiro:2004, Fu:2005, Svidzinsky:2008b, Svidzinsky:2010, Courteille:2010, Kuznetsov:2011, Bienaime:2012, Skipetrov:2014, Bellando:2014, Skipetrov:2015, Kuraptsev:2015, Guerin:2016a, Araujo:2016}.

In this work we use a version of the CD model formulated in the framework of the sequential quantum approach developed in \cite{Sokolov:2011}. Without repeating the derivation, let us note only its main features. We analyze the properties of a closed system consisting of all atoms and an electromagnetic field, including a vacuum reservoir. We look for the wave function $\psi$ of this system in the form of an expansion over the eigenfunctions ${\psi_l}$ of the Hamiltonian of noninteracting atoms and light $\psi = \sum_l b_l \psi_l$. Considering the case of weak excitation and restricting ourselves to the states of the atomic-field system containing no more than one photon, for the amplitudes $b_e$ of one-fold excited atomic states $\psi_e = | g \cdots e \cdots g \rangle$, we have the following set of equations:
\begin{equation}\label{e1}
\frac{\partial b_e}{\partial t} = \left( i\delta_e-\frac{\gamma}{2} \right)b_e -\frac{i\Omega_{e}}{2} + \frac{i\gamma}{2} \sum_{e' \neq e} V_{ee'}b_{e'}.
\end{equation}
Here the index $e$ denotes the excited atom in the state $\psi_e = | g \cdots e \cdots g \rangle$, as well as the specific populated Zeeman sublevel.

The first term on the right-hand side of Eq.\,(\ref{e1}) describes the natural evolution of independent atomic dipoles. The second one corresponds to the driving by the external laser field. The Rabi frequency of the field at the point where atom $e$ is located is $\Omega_e$. Its detuning $\delta_e$ may be different for different transitions $g\leftrightarrow e$.  The last term in Eq.\,(\ref{e1}) describes the pairwise dipole-dipole interaction and is consequently responsible for all polyatomic collective effects in the system. It reads
\begin{eqnarray}
V_{ee'}& =&
-\frac{2}{\gamma} \sum\limits_{\mu, \nu}
\mathbf{d}_{e g}^{\mu} \mathbf{d}_{g e'}^{\nu}
\frac{e^{i k r_{ij}}}{\hbar r_{ij}^3}
\nonumber
\\
&\times& \left\{
\vphantom{\frac{r_{ij}^{\mu} r_{ij}^{\nu}}{r_{ij}^2}}
 \delta_{\mu \nu}
\left[ 1 - i k r_{ij} - (k r_{ij})^2 \right]
\right.
 \\
&-&\left. \frac{\mathbf{r}_{ij}^{\mu} \mathbf{r}_{ij}^{\nu}}{r_{ij}^2}
\left[3 - 3 i k r_{ij} - (k r_{ij})^2 \right]
\right\}.
\nonumber
\end{eqnarray}
Here we assume that in the states $e$ and $e'$ atoms $i$ and $j$ are excited; $\mathbf{r}_{ij} =\mathbf{r}_i - \mathbf{r}_j$, $r_{ij} = |\mathbf{r}_i - \mathbf{r}_j|$, $\mathbf{d}_{e g}$ is the matrix element of the dipole moment operator for the transition $g \to e$, 
$k=\omega_0/c$ is the wave number associated with the transition, and $c$ is the speed of light in vacuum. The superscripts $\mu$ or $\nu$ denote projections of vectors on one of the axes $\mu$, $\nu = x$, $y$, and $z$ of the reference frame.

The system (\ref{e1}) is solved numerically many times for various random spatial configurations of motionless atoms. To do so we divide the entire temporal evolution into two stages: the stage of excitation of the ensemble by the external laser and the stage of free decay of the excitation, accompanied by fluorescence. The initial state for the first stage is chosen to be completely unexcited, when all atoms are in the ground state. In all calculations, the excitation is turned on at $t=-2000\tau_\at$ and ends at $t=0$. By this time, quasiequilibrium excitation of the ensemble is established, described by a set of amplitudes $b_e(t=0)$. These values are chosen as the initial condition for the second stage, at which we solved (\ref{e1}) with the laser switched off ($\Omega_e = 0$).

From the computed values of $b_{e}(t)$, we can find the amplitudes of all other states that determine the wave function $\psi$ of the joint atom-field system for a given configuration (for more details see \cite{Sokolov:2011}). Knowing the wave function, we can determine all the properties of both the atomic subsystem and the radiation.
For example, the intensity $I_\alpha (\boldsymbol{\Omega}, t)$ of the polarization component of the light $\alpha $, emitted by the cloud in a unit solid angle around an arbitrary direction given by the wave vector $\mathbf{k}$ ($\boldsymbol{\Omega} = {\theta, \varphi}$), can be defined as follows \cite{Kuraptsev:2017}:
\begin{equation}\label{e2}
I_\alpha(\boldsymbol{\Omega, t})=\frac{c}{4\pi} \left|  k^{2} \sum\limits_{i,m}\left(\mathbf{u}_\alpha^*\mathbf{d}_{ge}\right)  \beta_{e}(t) \exp \left(
-i\mathbf{k}\mathbf{r}_{i}\right) \right|^{2}.
\end{equation}
Here $\mathbf{u}_\alpha$ is the unit polarization vector of the secondary radiation.
It is also possible to compute the radiated field $E_\alpha(\mathbf{r}, t)$ and take the squared modulus of its Fourier transform to get the spectrum of the fluorescence, in a given time window.
The computed observables are averaged by the Monte Carlo method over the random spatial distribution of atoms.

\subsection{Parameters of the system}

Calculations of the dynamics of fluorescence by the CD method can be carried out for an arbitrary inhomogeneous spatial distribution of atoms and for an arbitrary shape of the atomic cloud. In our case, the shape and distribution have only a quantitative influence, without changing the basic physics of the investigated effects. For this reason, in this work, we consider the geometrically simplest case of an average homogeneous ensemble of cubic shape. Although the sharp boundaries might create specific modes \cite{Schilder:2016}, this choice simplifies the analysis of the spatial distribution of excited atoms and its change with time. In addition, for a simple geometry with relatively sharp boundaries of the atomic ensemble and a uniform spatial distribution of atoms (on average), it is possible to compare the numerical results with the predictions of the diffusion theory of radiative transfer.

The density of atoms $n$ in all calculations is the same: $n = 0.01 k^3$. Such a choice makes it possible to approximately simulate the dilute atomic ensembles that are dealt with in experiments. The excitation pulse is assumed to be rectangular; its carrier frequency $\omega$ is detuned from the transition frequency $\omega_0$ of a free atom by a detuning $\delta = \omega-\omega_0$. The pulse duration is $\gamma T = 2000$, which makes it possible to excite atoms in a rather narrow spectral range near the laser frequency. The laser radiation is assumed to be circularly polarized.

At a fixed density of atoms, a series of calculations was performed for different sizes $L$ of the atomic cloud and different detunings $\delta$. When choosing the detunings, we limited ourselves to the case when the optical thickness of the medium at the excitation frequency does not exceed a value on the order of 1. For the considered density and for the largest cloud containing $10^4$ atoms, it corresponds to $|\delta|\geq 2\gamma$.
We also note that all the following results were obtained for negative detunings. We have checked that the qualitative conclusions remain valid even with positive detunings, with only slight quantitative changes.

\section{Results}\label{sec:results}

\subsection{Decay dynamics of the fluorescence intensity}

First, we compute the total relative intensity $I(t)/I(0)$ of the fluorescence emitted in all directions and in all polarization channels. From that we compute the instant trapping time $\tau(t)$, which is defined as the inverse of the instantaneous decay rate $\tau(t) = 1/\Gamma(t)$, where
\begin{equation}
\Gamma(t) = - d\ln[I(t)]/dt.
\end{equation}

It was found in previous studies on subradiance \cite{Guerin:2016a, Araujo:2018} that the subradiant lifetime depends linearly on the resonant optical thickness $b_0=n\sigma_0 L$ ($\sigma_0=6\pi/k^2$ is the resonance cross section of a single atom). Here, our calculations reveal a more complex behavior of the fluorescence dynamics and its instantaneous decay. For each detuning $\delta$ there is a certain characteristic size of the atomic ensemble $L$ (or $b_0$), below which $I(t)/I(0)$ and $\tau(t)$ practically do not change with $b_0$.

\begin{figure}[tb]
\includegraphics[width=18.0pc]{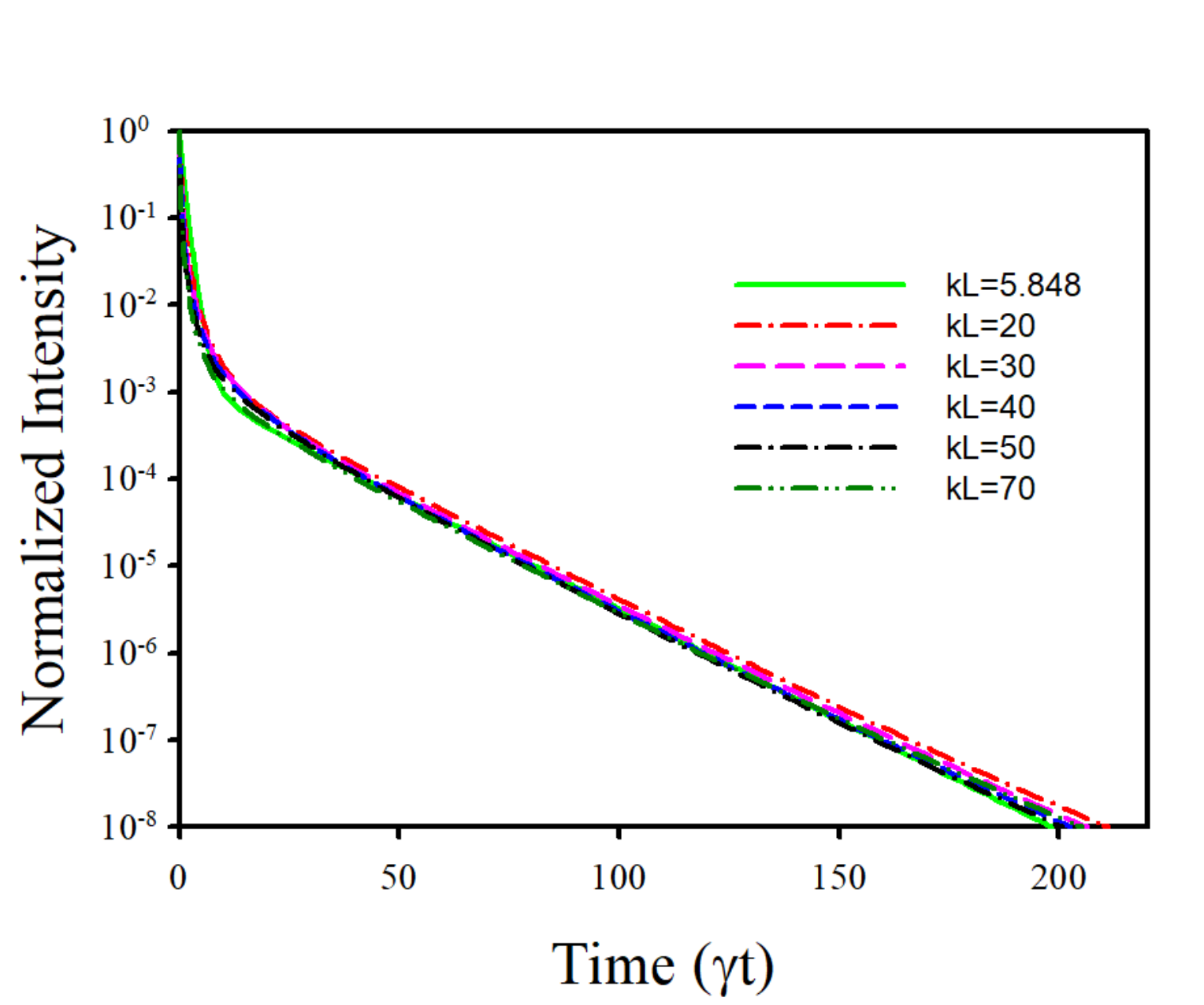}
\includegraphics[width=18.0pc]{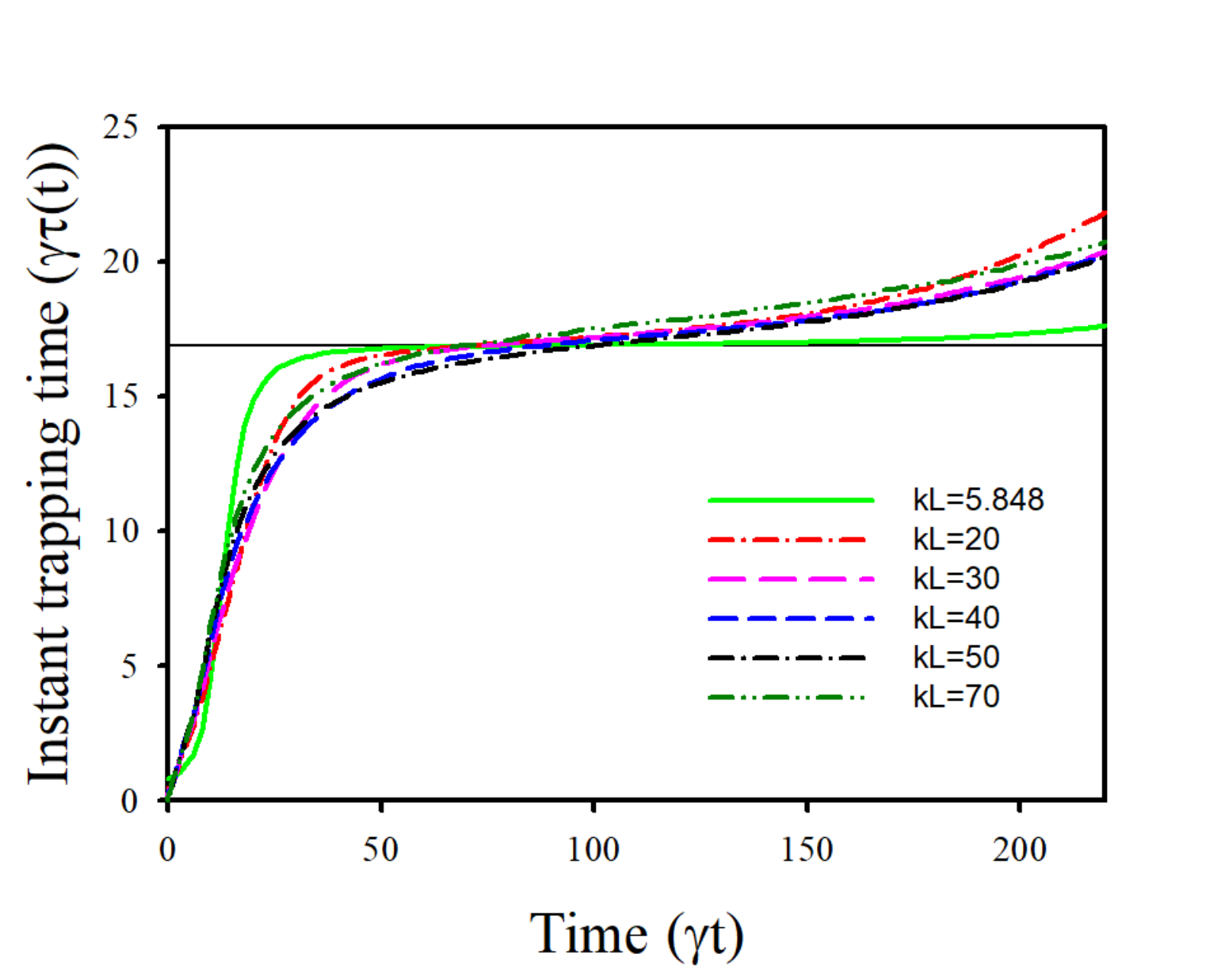}
\caption{Fluorescence dynamics of atomic ensembles of various sizes for (a) the total normalized radiation intensity in all directions and polarizations $I(t)/I(0)$ and (b) the instant trapping time $\tau(t)$. The density of atoms is $n = 0.01 k^3$, the duration of excitation is $\gamma T = 2000$, and the detuning is $\delta=-4\gamma$. The number of realizations changes from $10^4$ ($kL=20$) to 5000 ($kL=70$). The smallest size (green solid line) corresponds to $N=2$ atoms ($4\times10^6$ realizations in that case). The horizontal black solid line in (b) corresponds to the computed lifetime of subradiant dimers that are resonant with the excitation.}\label{f1}
\end{figure}

Figure \ref{f1} demonstrates such a behavior for $\delta=-4\gamma$. The curves correspond to different sizes (labeled in the figure), corresponding to different $b_0$ from $1.1$ to $13.2$ and different numbers of atoms from $N=2$ to $N=3430$.
Up to $kL=70$ the curves $I(t)/I(0)$ and $\tau(t)$ for different sizes of the ensemble practically do not differ within the accuracy of the calculations. In a significant range of time, they also do not differ much from the curves obtained for a diatomic ensemble $N=2$ (green solid line, $kL=5.848$). In the range from $\gamma t \simeq 50$ to $\gamma t \simeq 150$, the relative difference in the relative intensity between the results for the smallest number of atoms and the largest one does not exceed 5\%. In the middle of this time interval, it is significantly less, on the order of 1\%.

We attribute this behavior to the impact of pairs of close atoms (dimers) building subradiant states.
To support this interpretation, we compute
the level shifts $\Delta_r$ and the decay rate $\Gamma_r$ of the excited states of such diatomic quasimolecules, which depend on the distance $r$ between the atoms. For the considered case of $J_\mathrm{e}=1$, there are six states, two pairs of which are identical, with
\begin{eqnarray*}
\label{8}
\frac{\Delta_r}{\gamma}=\frac{3\epsilon}{4}\left( -\frac{p\cos(kr)}{kr}+q\left( \frac{\cos(kr)}{(kr)^3}+\frac{\sin(kr)}{(kr)^2} \right) \right),\\
\frac{\Gamma_r}{\gamma}=1-\frac{3\epsilon}{2}\left( -\frac{p\sin(kr)}{kr}+q\left( \frac{\sin(kr)}{(kr)^3}-\frac{\cos(kr)}{(kr)^2} \right) \right),
\end{eqnarray*}
where $\epsilon=\pm1$, $p_0=0$, $q_0=-2$, $p_{\pm 1}=1$, and $q_{\pm 1}=1$.

Among these states there are long-lived subradiant states. The energy shift corresponding to the detuning $\Delta_r=-4\gamma$ is achieved at $kr\approx 0.549$. At such a distance, the lifetime of the long-lived states is approximately equal to $1/\Gamma_r \approx 16.86\tau_\at$. The horizontal line in Fig.\,\ref{f1}(b) shows this value. It corresponds well to the observed lifetime in a large range of time and we have checked for several detunings that this correspondence works well. Long excitation by monochromatic light with this detuning is most likely to excite diatomic clusters with this distance between atoms, while other collective states are off-resonance. These clusters thus provide the main contribution to the fluorescence in a quite large time interval. At very late times, when these long-lived states have decayed, the main contribution to fluorescence come from longer-lived states and we observe an increase in the instant trapping time. Note that these pairs, whose influence on light transport in atomic gases was first discussed in \cite{Gero:2006}, were voluntarily removed in Refs. \cite{Guerin:2016a, Araujo:2018, Weiss:2018} by drawing the random atomic positions with an exclusion volume (see the discussion in Sec.\,\ref{sec:discuss}).

\begin{figure}[t]
\includegraphics[width=18.0pc]{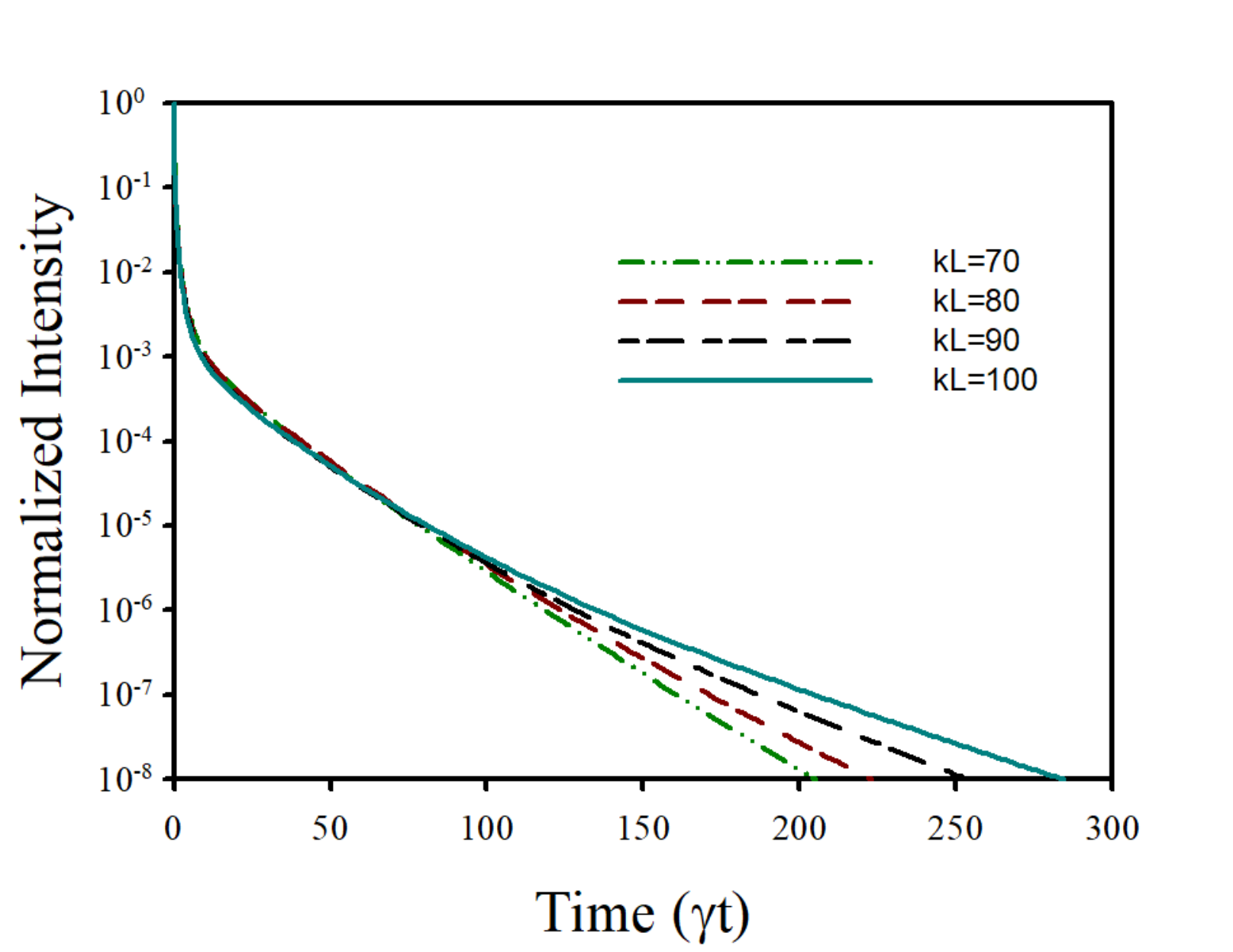}
\includegraphics[width=18.0pc]{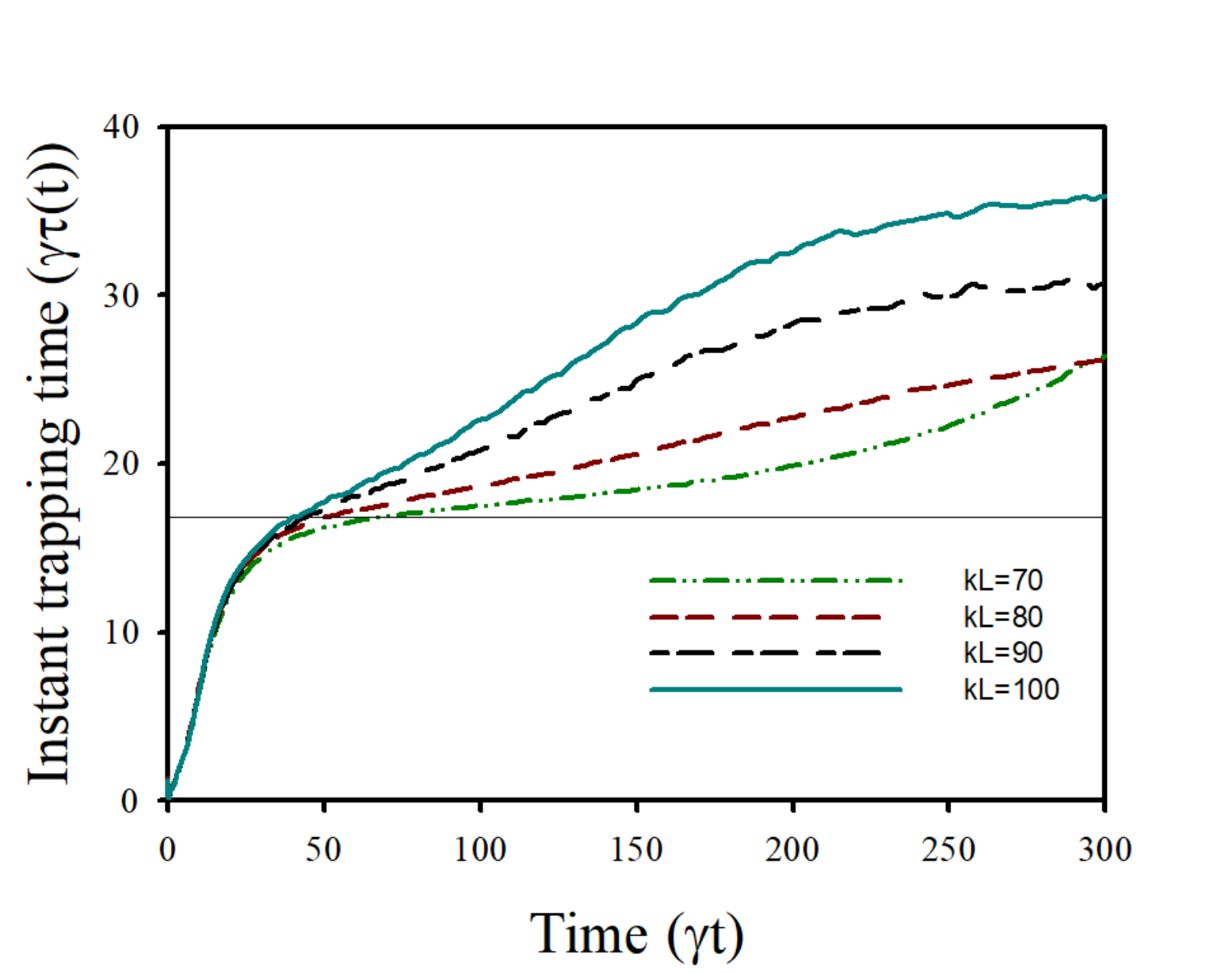}
\caption{Same as in Fig.\,\ref{f1} with larger sample sizes $L$. The number of realizations changes from 5000 ($kL=70$) to 300 ($kL=100$).}\label{f2}
\end{figure}

With a further increase of $b_0$, the situation changes (see Fig.\,\ref{f2}). After some time, the decay dynamics of the fluorescence begins to depend on the size of the atomic ensemble. For the parameters under consideration, such changes are observed for $\gamma t\geq 50$. For the sizes $kL\geq80$, one can clearly see an increase of $\tau(t)$, with a tendency towards stationary values, which depend on the size of the ensemble.

The difference in the behavior of the curves at long times in Figs.\,\ref{f1} and \ref{f2} shows that different mechanisms are at play. As discussed previously, for detuned excitation, moderate $b_0$ and moderate time delay, diatomic subradiance dominates. We argue in the following that, in the opposite regime of large $b_0$ and late time, slow diffusion of light due to incoherent multiple scattering, or radiation trapping, is the main contribution to the slow decay of the fluorescence \cite{Labeyrie:2003}.

Radiation trapping can only occur when the light mean free path is much shorter than the medium size, which is not the case for light detuned by several natural linewidths such as the incident field that we consider. However, even when the system is driven off-resonance, some resonant light is always present during the decay dynamics. This can be understood by the fact that the switching on and off of the driving field introduces some spectral broadening. Alternatively, in the coupled-dipole collective-mode picture, the presence of resonant light during the decay is due to the fact that, after the driving field is switched off, the weakly excited collective states decay at their own frequency, which is close to the natural atomic resonance. This generation of resonant light has been overlooked in previous papers discussing subradiance in dilute sample \cite{Bienaime:2012, Guerin:2016a, Weiss:2018}, which discarded a multiple-scattering interpretation of collective subradiance.

Radiation trapping introduces a new characteristic timescale, which is the lifetime of the longest diffusive mode
\begin{equation}\label{eq.tau_diff}
\tau_\diff=\frac{3b^2}{\alpha\pi^2}\tau_\at,
\end{equation}
where $\alpha$ depends on the shape of the atomic clouds \cite{Labeyrie:2003}. For homogeneous (on average) cubic ensembles, $\alpha=3$ \cite{Cao:2003}. This formula is valid only in the diffusive regime, i.e., when $b\gg1$. Also, the optical depth $b$ here differs slightly from $b_0$ by some addition $b'$ ($b=b_0+b'$) depending on the light mean free path. This difference is due to the extrapolation length for the boundary conditions of the radiative diffusion equation (see Fig.\,\ref{f4} and for more detail see, for example, \cite{Rossum:1999}).

It is possible to determine the parameters (size and detuning) for which radiation trapping begins to play a noticeable role by comparing this diffusive lifetime with the lifetime of long-lived states of diatomic quasimolecules $\tau_r=1/\Gamma_r$, the transition frequency of which coincides with the frequency of the exciting pulse $\Delta_r=\delta$. These are the quasi-molecules that are efficiently excited. When $\tau_\diff>\tau_r$, the main mechanism of subradiance should become radiation trapping after a certain time delay. Note, however, that after the decay of the slowest diffusion mode, the situation changes. Diatomic subradiance due to off-resonantly excited pairs, with smaller $r$, becomes dominant again. However, for large optical depths, this occurs at very late time after the end of the excitation, when the fluorescence signal becomes too small to be experimentally relevant.

\begin{figure}[tb]
\includegraphics[width=18pc]{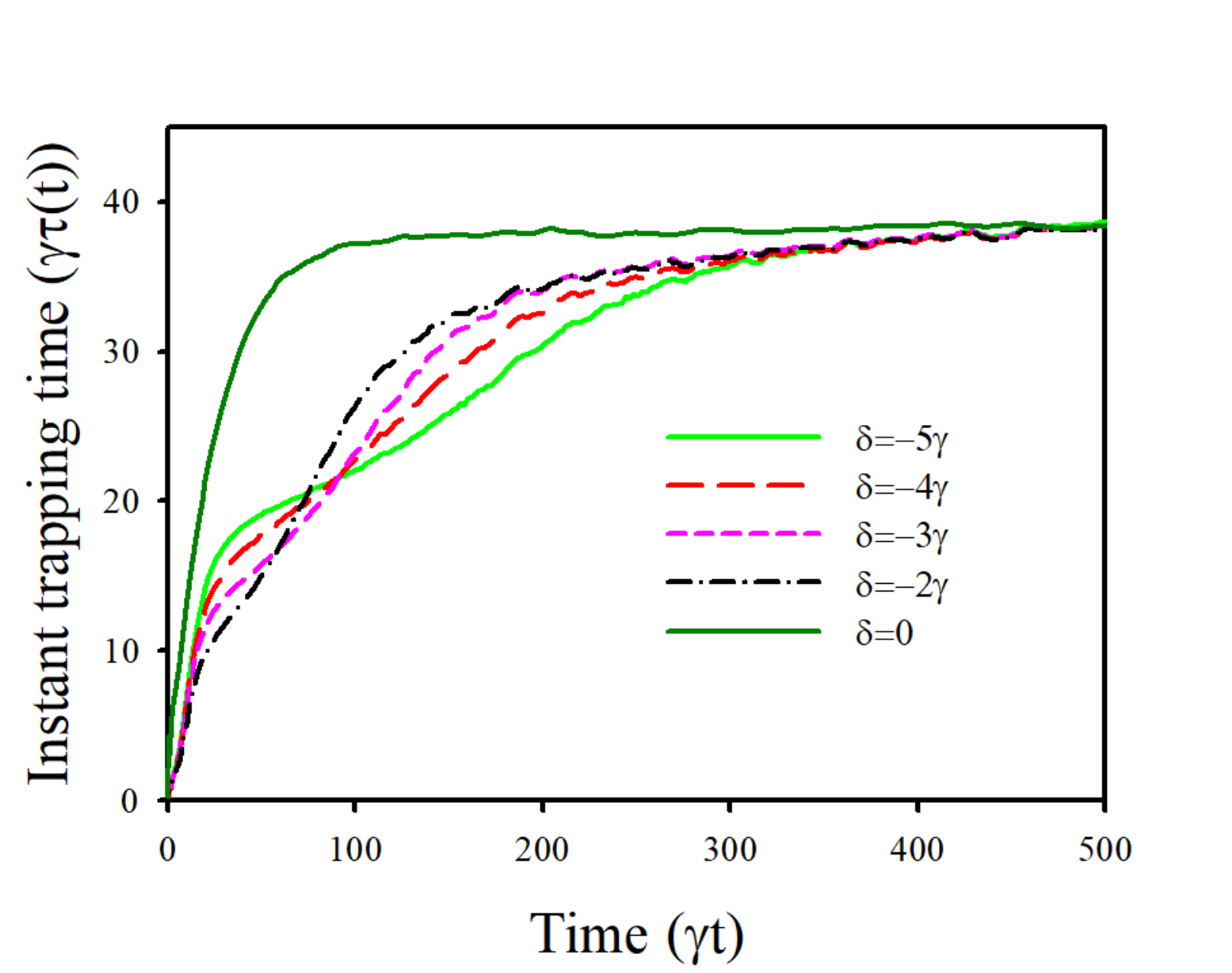}
	\caption{Instant trapping time $\tau(t)$ for different detunings $\delta$ of the exciting radiation. The system size is $kL=100$ ($b_0\approx 19$) and the number of realizations is 300. At late time radiation trapping dominates while at intermediate time and at large detuning, the decay rate due to diatomic clusters is clearly visible.}\label{f3}
\end{figure}

The comparison between $\tau_\diff$ and $\tau_r$, and consequently the respective roles of radiation trapping and diatomic subradiance on the fluorescence decay, depends not only on the size, but also on the frequency of the exciting light. The closer the frequency is to $\omega_0$, the more efficiently the diffuse mode is excited, the smaller $\tau_r$ is, and the smaller the relative role of diatomic clusters. This is well illustrated in Fig.\,\ref{f3}, computed with $b_0 \approx 19$ corresponding to $\tau_\diff \approx 36 \tau_\at$ (neglecting the extrapolation length): As the detuning $\delta$ increases, the time to reach the stationary decay rate (due to radiation trapping) also increases. The curves also show a tendency to reach an intermediate stationary value due to diatomic clusters, in the range around $\gamma t \sim 50$, which is more pronounced at large detuning (see the curve corresponding to $\delta=-5\gamma$).

In the following sections we present further arguments in favor of a diffusive interpretation of the slow decay at late time, by looking at other observables, namely, the spatial distribution of excitation, the spectrum and the polarization of the fluorescence.

\subsection{Spatial distribution of excitation}

\begin{figure}[b]
\includegraphics[width=18.0pc]{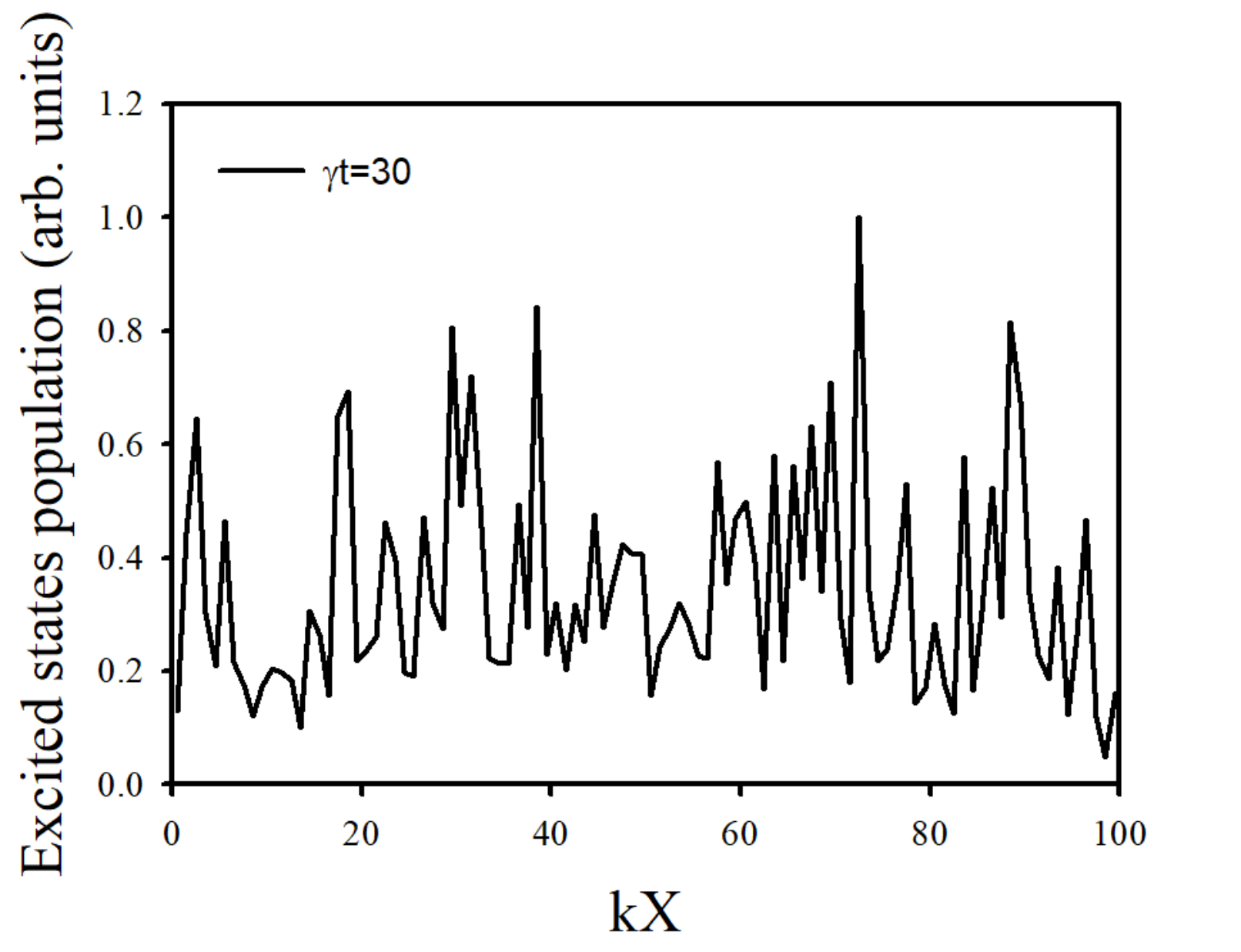}
\includegraphics[width=18.0pc]{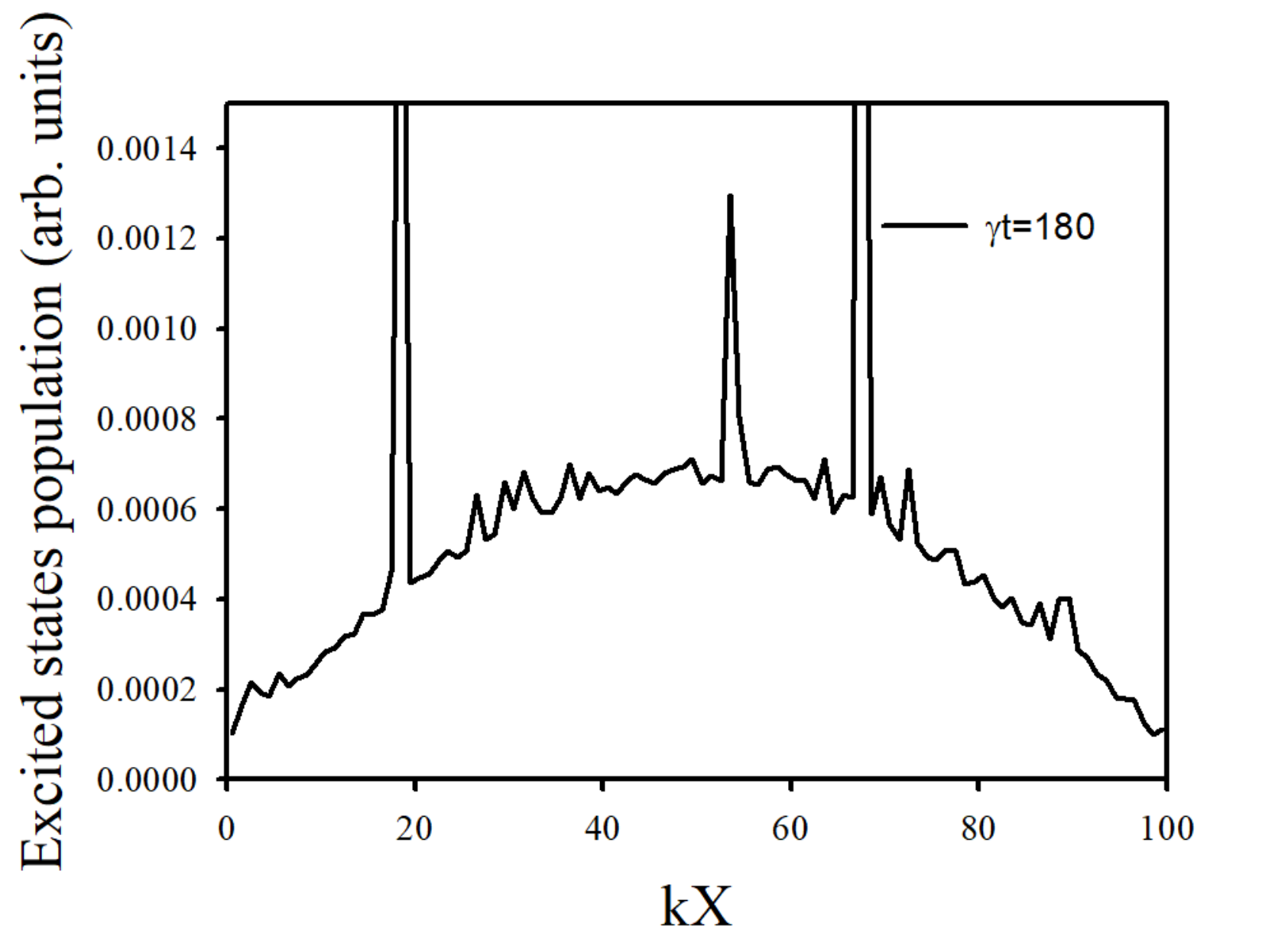}
\caption{Cut of the spatial distribution of excited atoms for two time delays after the switching off of the excitation: (a) $\gamma t=30$ and (b) $\gamma t=180$. The parameters are $kL=100$ and $\delta=-4\gamma$ and the number of realizations is 300. The width of the cut is $0.5/k$.}\label{f4}
\end{figure}

From the CD model we can analyze the spatial distribution of the excitation, which is encoded in $|b_e|^2$, and its change with time. This distribution is shown in Fig.\,\ref{f4} for two time delays after the end of the pulse.

At some intermediate stage of the slow fluorescence decay [$\gamma t = 30$, Fig.\,\ref{f4}(a)], the distribution of atomic excitation is practically homogeneous on average, although it exhibits relatively strong fluctuations, due to the limited number (300) of ensembles used for averaging, and the relatively small number of pairs. When these excited states decay, the main contribution begins to be made by the longer-lived diffusion mode, whose shape can be computed from the diffusion equation: It is characterized by a half-sine distribution over the ensemble [$\gamma t = 180$, Fig.\,\ref{f4}(b)]. At the boundaries of the cloud, the excitation is not zero. It vanishes at some distance on the order of the mean free path from the boundary of the cloud (extrapolation length). That is why the time $\tau_r$ [Eq.\,\ref{eq.tau_diff}] is determined not by the optical thickness $b_0$ but by a slightly larger value $b$.
Until the slowest diffusion mode decays, the distribution form does not change; only the absolute value of the excitation density decreases. There are also some localized spikes, which we associate with excited diatomic clusters (they each appear for only one particular realization of the disorder).

The transition from long-lived extended modes to longer-lived diffusive modes is consistent with the analysis of Ref.\,\cite{Guerin:2017b}, in which the relative population of the collective modes was studied as a function of the detuning of the driving field. It has been shown that the diffusive modes are very weakly populated when the system is driven out of resonance. That is why they are visible only after a long decay time, with correspondingly a very low relative fluorescence intensity. Note that the spatial profile of the driving beam also has an impact on which modes are preferentially populated \cite{Weiss:2018}.

\subsection{Fluorescence spectrum}

\begin{figure}[b]
\includegraphics[width=18.0pc]{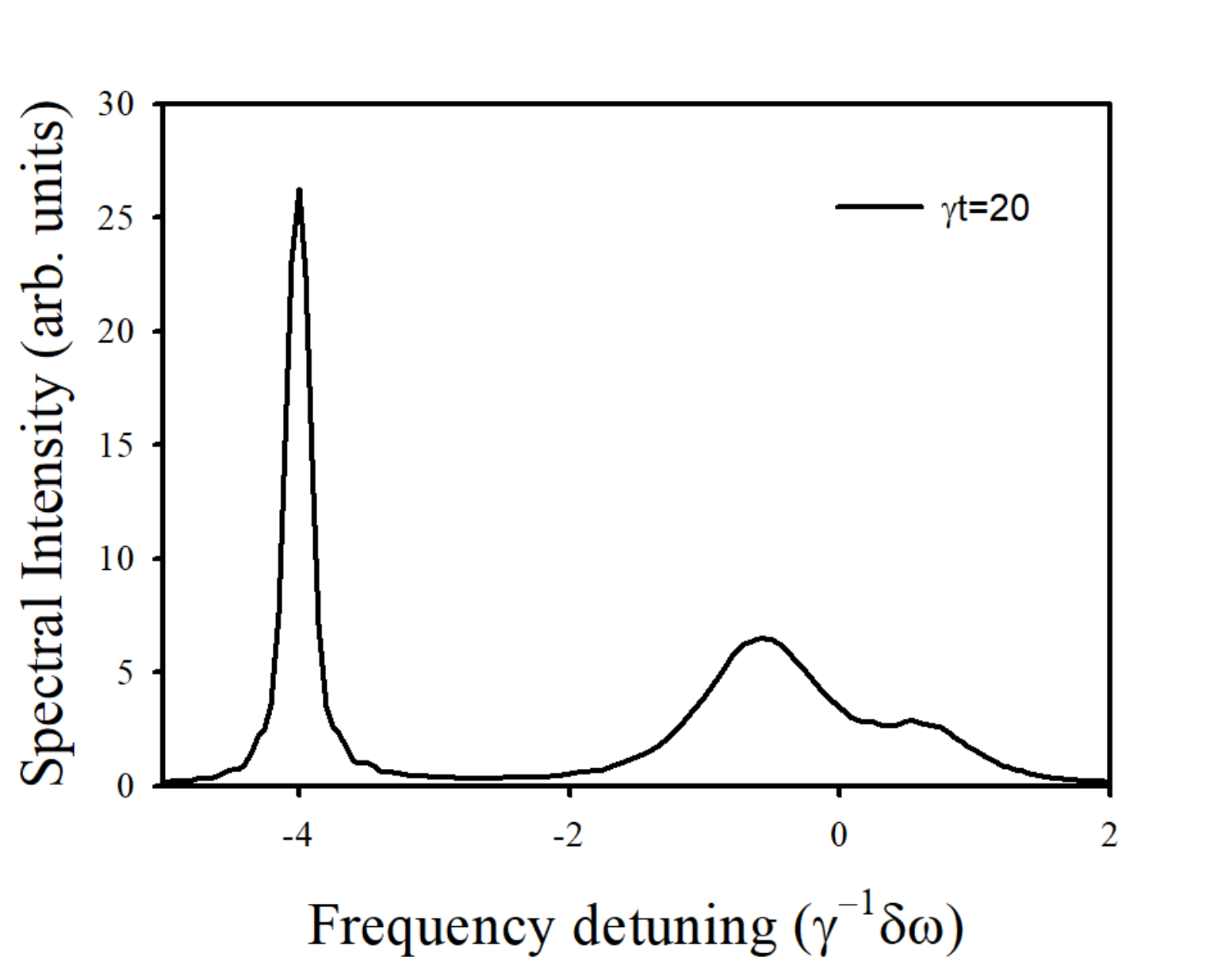}
\includegraphics[width=18.0pc]{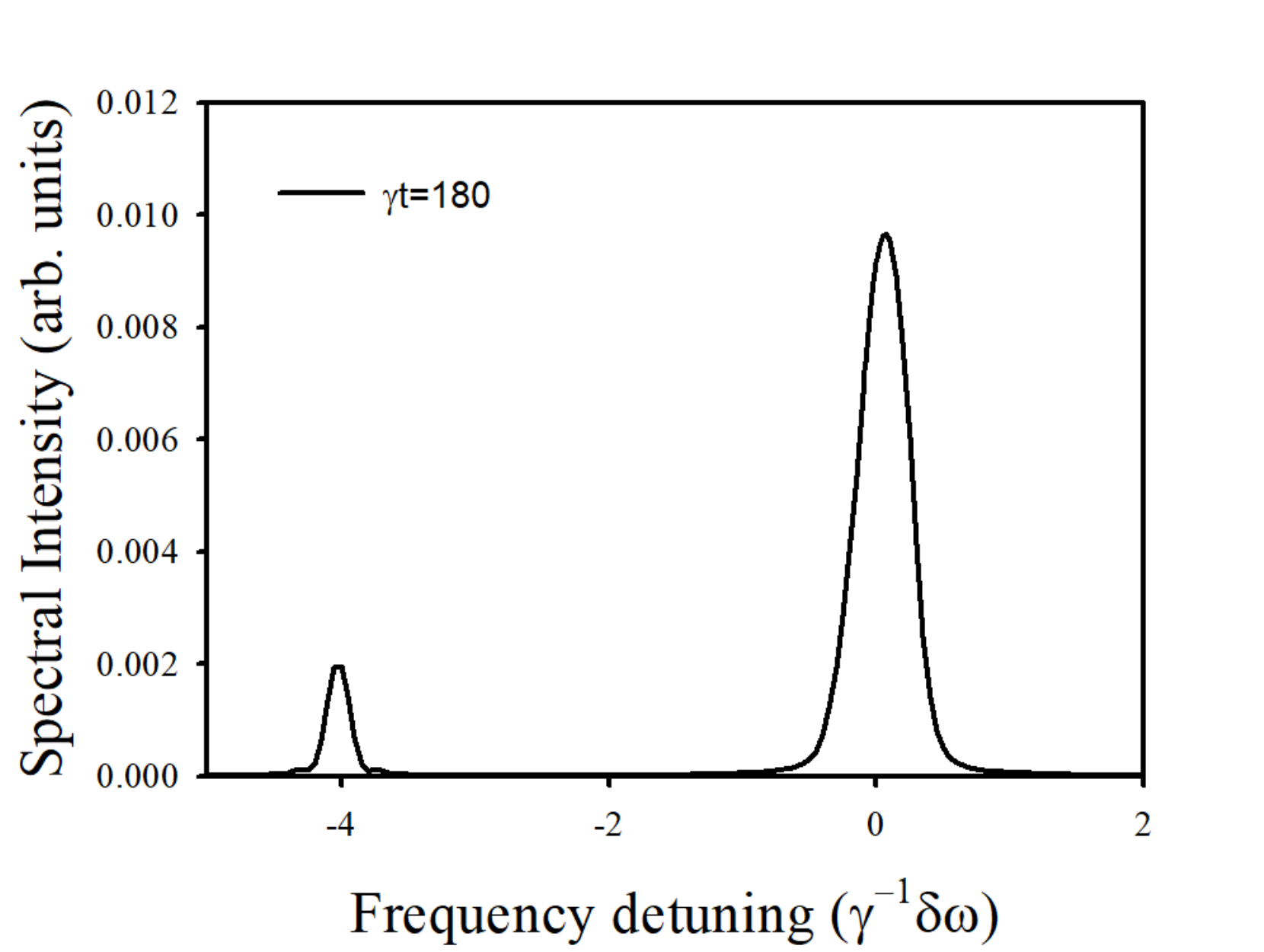}
\caption{Fluorescence spectrum for two time delays after the switching off of the excitation: (a) $\gamma t=20$ and (b) $\gamma t=180$. The parameters are $kL=100$ and $\delta=-4\gamma$.}\label{f5}
\end{figure}

The change of the decay mechanism with time can also be verified by analyzing the spectrum of the secondary radiation, which is determined by a short-term Fourier transform \cite{Bozhokin:2018} with a rectangular window of duration $ \gamma \Delta t = 30 $.

Figure \ref{f5}(a) shows the spectrum of the radiation scattered at an angle $\theta = \pi/4$ for the case when the middle of the time window corresponds to $\gamma t = 20$ after the end of the excitation. At this stage, we observe a Lorentzian line at the excitation frequency $\delta=-4\gamma$, as well as a distorted broad line near the free-atom resonance. This shape of the spectrum shows that the nonresonant driving field effectively excites diatomic quasimolecules, which are relatively few in a dilute medium, and with a low probability excites numerous collective states with frequencies close to the transition frequency of free atoms.

At later time the relative contribution of radiation at the excitation frequency decreases [$\gamma t = 180$, Fig.\,\ref{f5}(b)]. The main contribution to the radiation then comes from the collective states unshifted in frequency. This shows that at late time, the emitted radiation is mainly at the atomic resonance and thus can undergo multiple scattering in the optically thick medium.

\subsection{Fluorescence polarization}

\begin{figure}[b]
\includegraphics[width=17.0pc]{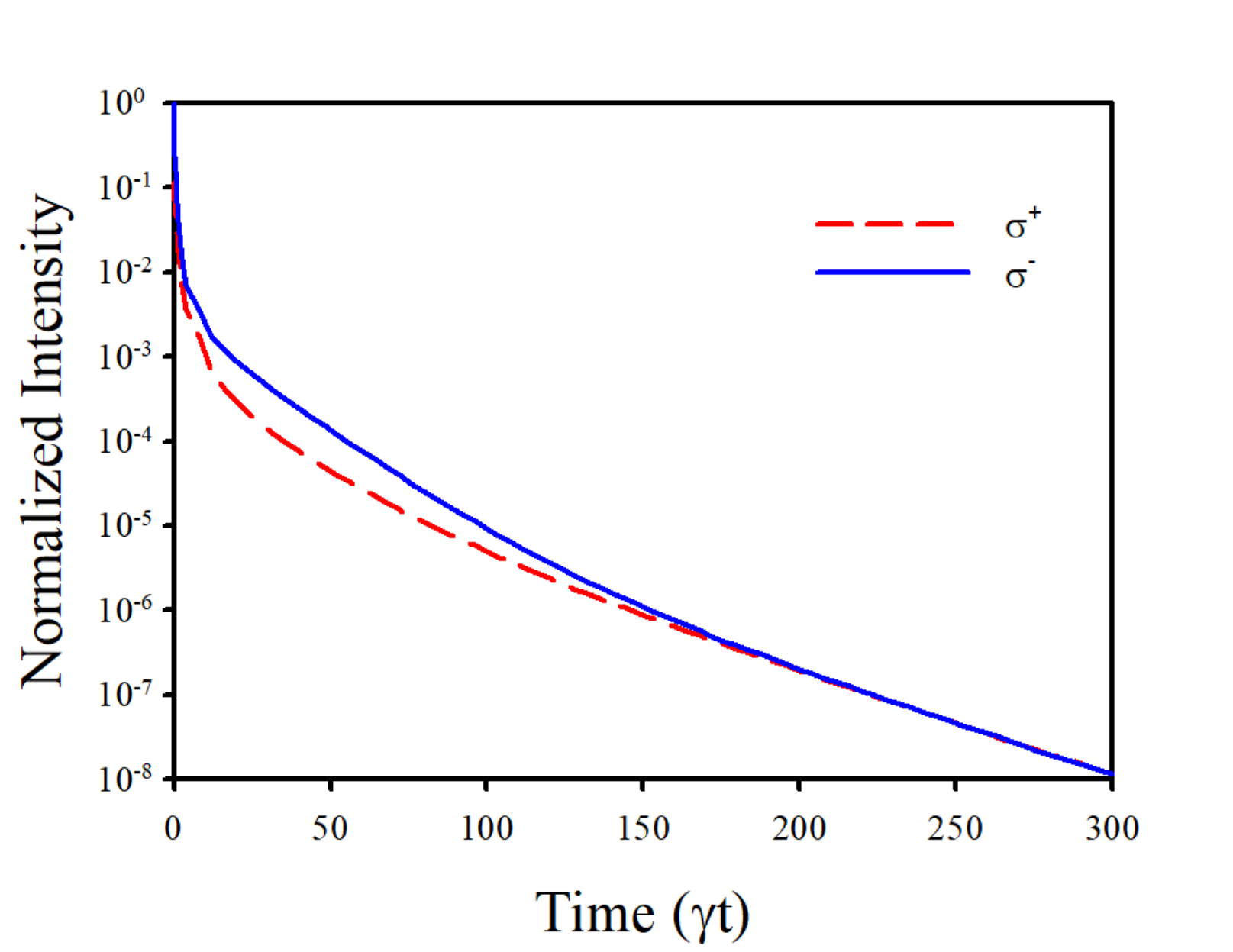}
\includegraphics[width=17.0pc]{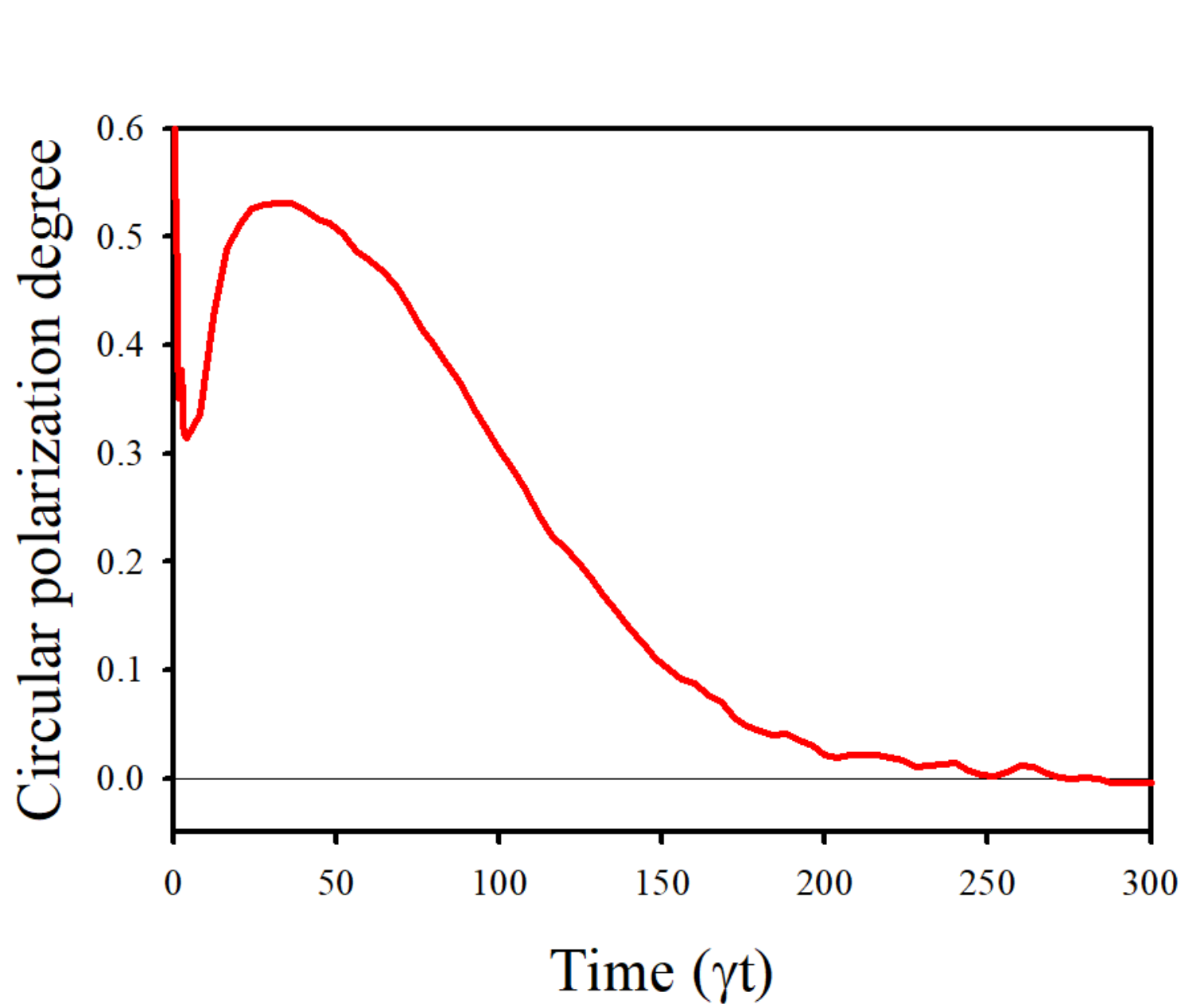}
\caption{(a) Normalized intensities of two orthogonal circularly polarized components of the radiation emitted at an angle $\theta = \pi/4$. The incident radiation is left circularly polarized and the other parameters are $kL=100$ and $\delta=-4\gamma$. (b) Corresponding degree of circular polarization.}\label{f6}
\end{figure}

Another interesting property of the slow-decaying fluorescence is its polarization. In Fig.\,\ref{f6} we show the time dependence of the decaying intensities in two orthogonal polarization channels. We observe that at very short time, during the superradiant decay, light is polarized, which is consistent with a single-scattering interpretation of superradiance, such as the one given in \cite{Weiss:2021}. Then when multiple scattering starts to occur, the degree of polarization decreases. At intermediate time though, light remains partially polarized, which is consistent with radiation by diatomic clusters. Finally, at later time ($\gamma t \gtrsim 150$), light is completely depolarized, with equal intensities in the two polarization channels, which is consistent with multiply scattered light. This observation, also reported in \cite{Cipris:2021b}, supports the interpretation that late-time subradiance is due to radiation trapping.

\section{Discussion and conclusion}\label{sec:discuss}

In this work we have discussed the nature of subradiance in dilute samples. Indeed, different optical phenomena can be responsible for long-lived collective modes.

We found that, if the system is driven off resonance, subradiance between pairs of very close atoms may have a significant contribution, even though the density of the sample is low on average. This is because the pairs build collective modes with a large energy shift that can compensate for the laser detuning, such that the corresponding modes are particularly well coupled to the driving field. This regime is characterized by a decay rate mainly related to the detuning and hardly affected by the parameters of the sample. However, this is valid within the idealized conditions of this work, which neglects atomic motion and other mechanical effects such as collisions. It is plausible that in real experimental conditions, the influence of these pairs is suppressed by thermal motions or by the attractive force induced by the near-field interaction and the subsequent inelastic collision and light emission \cite{Caires:2004, Fuhrmanek:2012}. That is why previous works used an exclusion volume to remove the pairs \cite{Guerin:2016a, Araujo:2018, Weiss:2018, Cipris:2021b}. This is consistent with the fact that experimental results never exhibit lifetimes that are independent of $b_0$, as shown in Fig.\,\ref{f1}. Nevertheless, further numerical studies, using moving atoms \cite{Weiss:2019, Kuraptsev:2020} and possibly including collective forces \cite{Gisbert:2019}, would be desirable to conclude on the possible role of pairs in experiments.

When pair subradiance does not dominate (for larger resonant optical depth or later time), we have found that the slow decay was consistent with radiation trapping for the following properties: emitted polarization, emitted spectrum and intensity distribution inside the sample. This multiple-scattering interpretation of collective subradiance was discarded in some previous works \cite{Bienaime:2012, Guerin:2016a, Weiss:2018} because it counter-intuitively also happens when the system is driven off-resonance. However, we have shown here that some resonant light is always generated at the switching off; therefore, multiple scattering is a valid interpretation for the long-lived modes.

Nevertheless, one effect that does not fully support the radiation-trapping interpretation is the scaling of subradiant lifetimes with the optical depth, which has been repeatedly found to be linear \cite{Bienaime:2012, Guerin:2016a, Araujo:2018, Weiss:2018, Weiss:2019, Cipris:2021b}, while diffusion theory predicts a quadratic scaling [Eq.\,(\ref{eq.tau_diff})]. This fact is still not well understood. In experiments, atomic motion is known to break the quadratic dependence because of the Doppler-induced frequency redistribution \cite{Labeyrie:2003, Labeyrie:2005, Pierrat:2009}. In the CD simulations at zero temperature, the quadratic dependence should be reached for long-enough time and large-enough optical thickness, if diffusion theory is valid. We have not been able to observe this quadratic scaling so far.

\begin{figure}[t]
\includegraphics[width=18.0pc]{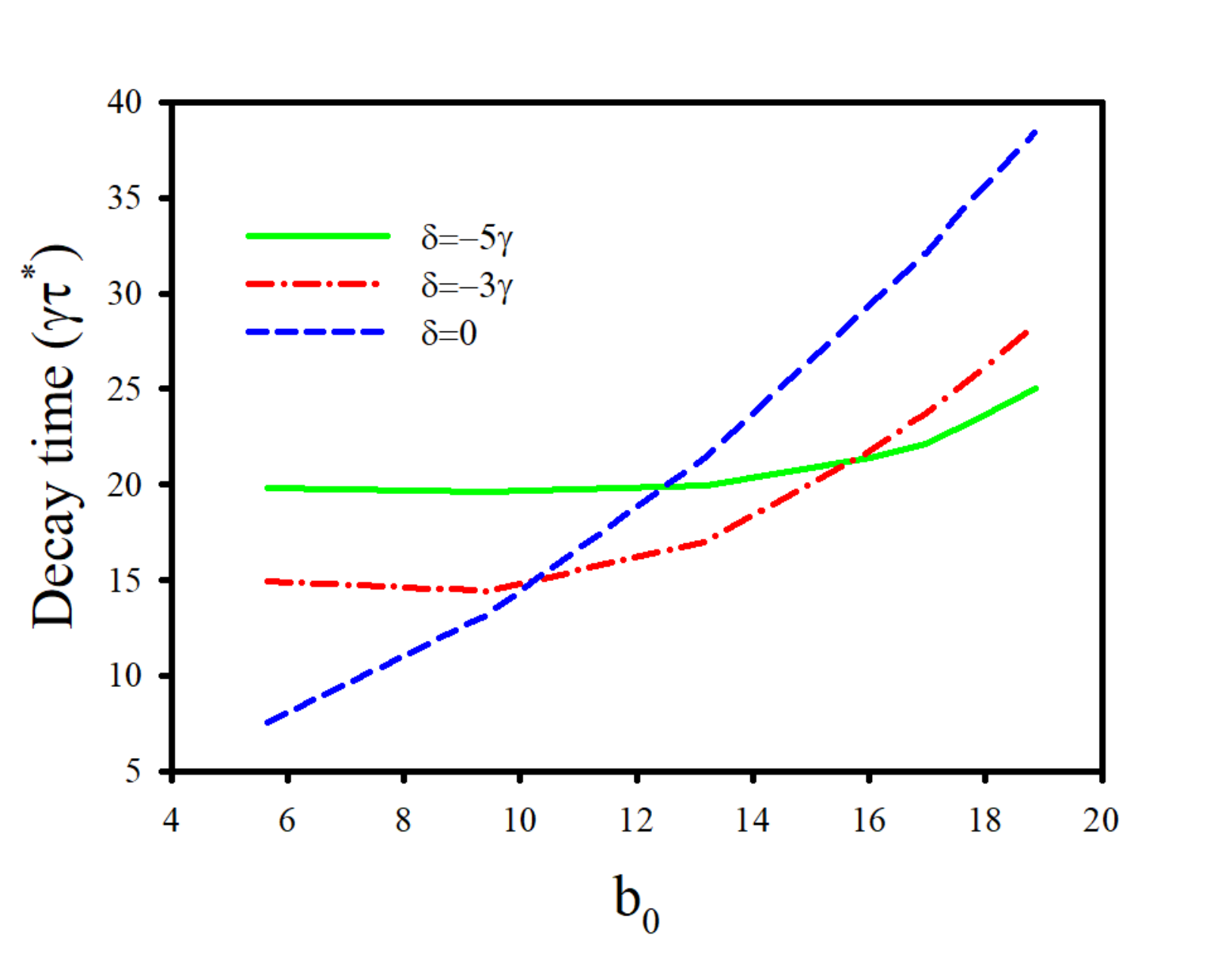}
\caption{Dependence of the characteristic decay time $\tau^*$ with the resonant optical thickness for different detuning $\delta$ of the driving field.}\label{f7}
\end{figure}

This might be due to the computational limitations of the CD model: Having large-enough $b_0$ to reach the diffusive regime without introducing high-density effects \cite{Cipris:2021b} needs atom numbers that are out of reach, or at least very difficult to handle.
To illustrate those difficulties we show in Fig.\,\ref{f7} the instant trapping time $\tau^*$ when the fluorescence intensity is $10^{-6}$ from the initial one, a value compatible with experimental observations, as a function of $b_0$, up to $b_0 \approx 19$ ($N=10000$). For detuned excitation, there is a range of $b_0$ for which the decay time remains practically constant. This constant value increases with the detuning. In this range of parameters, the dominant contribution to the fluorescence is made by diatomic clusters. At larger optical depth, the effect of radiation trapping is visible and the decay time increases. However, we do not reach the regime where a quadratic scaling would be clearly visible. We have checked that similar results are obtained if the fluorescence level is chosen at $10^{-8}$. 

Another possible reason for not seeing the quadratic scaling might be the breakdown of a diffusion approach, which practically neglects all wave effects. Therefore, it cannot be excluded that the richer physics included in the CD model and in experiments changes the behavior of the long-lived modes. This could be tested by a careful comparison between the results provided by the CD model and by a random walk model, as previously used for describing radiation trapping \cite{Labeyrie:2003, Labeyrie:2005, Datsyuk:2006, Pierrat:2009, Weiss:2018}. However, the two models have different computational limitations such that a direct comparison is challenging for dynamical problems (it has been done for steady-state scattering in \cite{Chabe:2014,Sokolov:2019}). It is thus left for future work.

Finally, we note that the numerical methods developed to study the polarization and spectrum from the CD approach can also be used in other contexts, such as highly sensitive laser polarization optical analysis of the properties of ensembles of impurities in perfect optical crystals \cite{Fofanov:2020}.

\section*{Acknowledgments}

We thank A.~Cipris, P.~Weiss, and R.~Bachelard for fruitful discussions. The work was carried out within the framework of the State Task for Basic Research (topic code FSEG-2020-0024). Part of this work was performed in the framework of State Task 075-00280-21-00 on the topic No. 0074-2019-0007 of the Ministry of Science and Higher Education of the Russian Federation and other parts in the framework of the European project ANDLICA, ERC Advanced Grant No.\,832219. We also acknowledge funding from the French National Research Agency (Projects PACE-IN No. ANR19-QUAN-003 and QuaCor No. ANR19-CE47-0014).


\begin{thebibliography}{63}%
\makeatletter
\providecommand \@ifxundefined [1]{%
 \@ifx{#1\undefined}
}%
\providecommand \@ifnum [1]{%
 \ifnum #1\expandafter \@firstoftwo
 \else \expandafter \@secondoftwo
 \fi
}%
\providecommand \@ifx [1]{%
 \ifx #1\expandafter \@firstoftwo
 \else \expandafter \@secondoftwo
 \fi
}%
\providecommand \natexlab [1]{#1}%
\providecommand \enquote  [1]{``#1''}%
\providecommand \bibnamefont  [1]{#1}%
\providecommand \bibfnamefont [1]{#1}%
\providecommand \citenamefont [1]{#1}%
\providecommand \href@noop [0]{\@secondoftwo}%
\providecommand \href [0]{\begingroup \@sanitize@url \@href}%
\providecommand \@href[1]{\@@startlink{#1}\@@href}%
\providecommand \@@href[1]{\endgroup#1\@@endlink}%
\providecommand \@sanitize@url [0]{\catcode `\\12\catcode `\$12\catcode
  `\&12\catcode `\#12\catcode `\^12\catcode `\_12\catcode `\%12\relax}%
\providecommand \@@startlink[1]{}%
\providecommand \@@endlink[0]{}%
\providecommand \url  [0]{\begingroup\@sanitize@url \@url }%
\providecommand \@url [1]{\endgroup\@href {#1}{\urlprefix }}%
\providecommand \urlprefix  [0]{URL }%
\providecommand \Eprint [0]{\href }%
\providecommand \doibase [0]{http://dx.doi.org/}%
\providecommand \selectlanguage [0]{\@gobble}%
\providecommand \bibinfo  [0]{\@secondoftwo}%
\providecommand \bibfield  [0]{\@secondoftwo}%
\providecommand \translation [1]{[#1]}%
\providecommand \BibitemOpen [0]{}%
\providecommand \bibitemStop [0]{}%
\providecommand \bibitemNoStop [0]{.\EOS\space}%
\providecommand \EOS [0]{\spacefactor3000\relax}%
\providecommand \BibitemShut  [1]{\csname bibitem#1\endcsname}%
\let\auto@bib@innerbib\@empty
\bibitem [{\citenamefont {Chang}\ \emph {et~al.}(2018)\citenamefont {Chang},
  \citenamefont {Douglas}, \citenamefont {Gonz\'ales-Tudela}, \citenamefont
  {Hung},\ and\ \citenamefont {Kimble}}]{Chang:2018}%
  \BibitemOpen
  \bibfield  {author} {\bibinfo {author} {\bibfnamefont {D.~E.}\ \bibnamefont
  {Chang}}, \bibinfo {author} {\bibfnamefont {J.~S.}\ \bibnamefont {Douglas}},
  \bibinfo {author} {\bibfnamefont {A.}~\bibnamefont {Gonz\'ales-Tudela}},
  \bibinfo {author} {\bibfnamefont {C.-L.}\ \bibnamefont {Hung}}, \ and\
  \bibinfo {author} {\bibfnamefont {H.~J.}\ \bibnamefont {Kimble}},\ }\bibfield
   {title} {\enquote {\bibinfo {title} {\textit{Colloquium}: Quantum matter
  built from nanoscopic lattices of atoms and photons},}\ }\href@noop {}
  {\bibfield  {journal} {\bibinfo  {journal} {Rev. Mod. Phys.}\ }\textbf
  {\bibinfo {volume} {90}},\ \bibinfo {pages} {031002} (\bibinfo {year}
  {2018})}\BibitemShut {NoStop}%
\bibitem [{\citenamefont {Dicke}(1954)}]{Dicke:1954}%
  \BibitemOpen
  \bibfield  {author} {\bibinfo {author} {\bibfnamefont {R.~H.}\ \bibnamefont
  {Dicke}},\ }\bibfield  {title} {\enquote {\bibinfo {title} {Coherence in
  spontaneous radiation processes},}\ }\href@noop {} {\bibfield  {journal}
  {\bibinfo  {journal} {Phys. Rev.}\ }\textbf {\bibinfo {volume} {93}},\
  \bibinfo {pages} {99--110} (\bibinfo {year} {1954})}\BibitemShut {NoStop}%
\bibitem [{\citenamefont {Pavolini}\ \emph {et~al.}(1985)\citenamefont
  {Pavolini}, \citenamefont {Crubellier}, \citenamefont {Pillet}, \citenamefont
  {Cabaret},\ and\ \citenamefont {Liberman}}]{Pavolini:1985}%
  \BibitemOpen
  \bibfield  {author} {\bibinfo {author} {\bibfnamefont {D.}~\bibnamefont
  {Pavolini}}, \bibinfo {author} {\bibfnamefont {A.}~\bibnamefont
  {Crubellier}}, \bibinfo {author} {\bibfnamefont {P.}~\bibnamefont {Pillet}},
  \bibinfo {author} {\bibfnamefont {L.}~\bibnamefont {Cabaret}}, \ and\
  \bibinfo {author} {\bibfnamefont {S.}~\bibnamefont {Liberman}},\ }\bibfield
  {title} {\enquote {\bibinfo {title} {Experimental evidence for
  subradiance},}\ }\href@noop {} {\bibfield  {journal} {\bibinfo  {journal}
  {Phys. Rev. Lett.}\ }\textbf {\bibinfo {volume} {54}},\ \bibinfo {pages}
  {1917--1920} (\bibinfo {year} {1985})}\BibitemShut {NoStop}%
\bibitem [{\citenamefont {Bienaim\'e}\ \emph {et~al.}(2012)\citenamefont
  {Bienaim\'e}, \citenamefont {Piovella},\ and\ \citenamefont
  {Kaiser}}]{Bienaime:2012}%
  \BibitemOpen
  \bibfield  {author} {\bibinfo {author} {\bibfnamefont {T.}~\bibnamefont
  {Bienaim\'e}}, \bibinfo {author} {\bibfnamefont {N.}~\bibnamefont
  {Piovella}}, \ and\ \bibinfo {author} {\bibfnamefont {R.}~\bibnamefont
  {Kaiser}},\ }\bibfield  {title} {\enquote {\bibinfo {title} {Controlled
  {Dicke} subradiance from a large cloud of two-level systems},}\ }\href@noop
  {} {\bibfield  {journal} {\bibinfo  {journal} {Phys. Rev. Lett.}\ }\textbf
  {\bibinfo {volume} {108}},\ \bibinfo {pages} {123602} (\bibinfo {year}
  {2012})}\BibitemShut {NoStop}%
\bibitem [{\citenamefont {Guerin}\ \emph {et~al.}(2016)\citenamefont {Guerin},
  \citenamefont {Ara\'ujo},\ and\ \citenamefont {Kaiser}}]{Guerin:2016a}%
  \BibitemOpen
  \bibfield  {author} {\bibinfo {author} {\bibfnamefont {W.}~\bibnamefont
  {Guerin}}, \bibinfo {author} {\bibfnamefont {M.~O.}\ \bibnamefont
  {Ara\'ujo}}, \ and\ \bibinfo {author} {\bibfnamefont {R.}~\bibnamefont
  {Kaiser}},\ }\bibfield  {title} {\enquote {\bibinfo {title} {Subradiance in a
  large cloud of cold atoms},}\ }\href@noop {} {\bibfield  {journal} {\bibinfo
  {journal} {Phys. Rev. Lett.}\ }\textbf {\bibinfo {volume} {116}},\ \bibinfo
  {pages} {083601} (\bibinfo {year} {2016})}\BibitemShut {NoStop}%
\bibitem [{\citenamefont {Jenkins}\ \emph {et~al.}(2017)\citenamefont
  {Jenkins}, \citenamefont {Ruostekoski}, \citenamefont {Papasimakis},
  \citenamefont {Savo},\ and\ \citenamefont {Zheludev}}]{Jenkins2017}%
  \BibitemOpen
  \bibfield  {author} {\bibinfo {author} {\bibfnamefont {S.~D.}\ \bibnamefont
  {Jenkins}}, \bibinfo {author} {\bibfnamefont {J.}~\bibnamefont
  {Ruostekoski}}, \bibinfo {author} {\bibfnamefont {N.}~\bibnamefont
  {Papasimakis}}, \bibinfo {author} {\bibfnamefont {S.}~\bibnamefont {Savo}}, \
  and\ \bibinfo {author} {\bibfnamefont {N.~I.}\ \bibnamefont {Zheludev}},\
  }\bibfield  {title} {\enquote {\bibinfo {title} {Many-body subradiant
  excitations in metamaterial arrays: Experiment and theory},}\ }\href
  {\doibase 10.1103/PhysRevLett.119.053901} {\bibfield  {journal} {\bibinfo
  {journal} {Phys. Rev. Lett.}\ }\textbf {\bibinfo {volume} {119}},\ \bibinfo
  {pages} {053901} (\bibinfo {year} {2017})}\BibitemShut {NoStop}%
\bibitem [{\citenamefont {Solano}\ \emph {et~al.}(2017)\citenamefont {Solano},
  \citenamefont {Barberis-Blostein}, \citenamefont {Fatemi}, \citenamefont
  {Orozco},\ and\ \citenamefont {Rolston}}]{Solano:2017}%
  \BibitemOpen
  \bibfield  {author} {\bibinfo {author} {\bibfnamefont {P.}~\bibnamefont
  {Solano}}, \bibinfo {author} {\bibfnamefont {P.}~\bibnamefont
  {Barberis-Blostein}}, \bibinfo {author} {\bibfnamefont {F.~K.}\ \bibnamefont
  {Fatemi}}, \bibinfo {author} {\bibfnamefont {L.~A.}\ \bibnamefont {Orozco}},
  \ and\ \bibinfo {author} {\bibfnamefont {S.~L.}\ \bibnamefont {Rolston}},\
  }\bibfield  {title} {\enquote {\bibinfo {title} {Super-radiance reveals
  infinite-range dipole interactions through a nanofiber},}\ }\href@noop {}
  {\bibfield  {journal} {\bibinfo  {journal} {Nature Comm.}\ }\textbf {\bibinfo
  {volume} {8}},\ \bibinfo {pages} {1857} (\bibinfo {year} {2017})}\BibitemShut
  {NoStop}%
\bibitem [{\citenamefont {Rui}\ \emph {et~al.}(2020)\citenamefont {Rui},
  \citenamefont {Wei}, \citenamefont {{Rubio-Abadal}}, \citenamefont
  {Hollerith}, \citenamefont {Zeiher}, \citenamefont {{Stamper-Kurn}},
  \citenamefont {Gross},\ and\ \citenamefont {Bloch}}]{Rui:2020}%
  \BibitemOpen
  \bibfield  {author} {\bibinfo {author} {\bibfnamefont {J.}~\bibnamefont
  {Rui}}, \bibinfo {author} {\bibfnamefont {D.}~\bibnamefont {Wei}}, \bibinfo
  {author} {\bibfnamefont {A.}~\bibnamefont {{Rubio-Abadal}}}, \bibinfo
  {author} {\bibfnamefont {S.}~\bibnamefont {Hollerith}}, \bibinfo {author}
  {\bibfnamefont {J.}~\bibnamefont {Zeiher}}, \bibinfo {author} {\bibfnamefont
  {D.~M.}\ \bibnamefont {{Stamper-Kurn}}}, \bibinfo {author} {\bibfnamefont
  {C.}~\bibnamefont {Gross}}, \ and\ \bibinfo {author} {\bibfnamefont
  {I.}~\bibnamefont {Bloch}},\ }\bibfield  {title} {\enquote {\bibinfo {title}
  {A subradiant optical mirror formed by a single structured atomic layer},}\
  }\href@noop {} {\bibfield  {journal} {\bibinfo  {journal} {Nature}\ }\textbf
  {\bibinfo {volume} {583}},\ \bibinfo {pages} {369--375} (\bibinfo {year}
  {2020})}\BibitemShut {NoStop}%
\bibitem [{\citenamefont {Ferioli}\ \emph {et~al.}(2020)\citenamefont
  {Ferioli}, \citenamefont {Glicenstein}, \citenamefont {Henriet},
  \citenamefont {{Ferrier-Barbut}},\ and\ \citenamefont
  {Browaeys}}]{Ferioli:2021}%
  \BibitemOpen
  \bibfield  {author} {\bibinfo {author} {\bibfnamefont {G.}~\bibnamefont
  {Ferioli}}, \bibinfo {author} {\bibfnamefont {A.}~\bibnamefont
  {Glicenstein}}, \bibinfo {author} {\bibfnamefont {L.}~\bibnamefont
  {Henriet}}, \bibinfo {author} {\bibfnamefont {I.}~\bibnamefont
  {{Ferrier-Barbut}}}, \ and\ \bibinfo {author} {\bibfnamefont
  {A.}~\bibnamefont {Browaeys}},\ }\href@noop {} {\enquote {\bibinfo {title}
  {Storage and release of subradiant excitations in a dense atomic cloud},}\ }
  \href {\doibase
  10.1103/PhysRevX.11.021031} {\bibfield  {journal} {\bibinfo  {journal}
  {Phys. Rev. X}\ }\textbf {\bibinfo {volume} {11}},\ \bibinfo {pages}
  {021031} (\bibinfo {year} {2021})}\BibitemShut {NoStop}%
\bibitem [{\citenamefont {Scully}(2015)}]{Scully:2015}%
  \BibitemOpen
  \bibfield  {author} {\bibinfo {author} {\bibfnamefont {M.~O.}\ \bibnamefont
  {Scully}},\ }\bibfield  {title} {\enquote {\bibinfo {title} {Single photon
  subradiance: Quantum control of spontaneous emission and ultrafast
  readout},}\ }\href@noop {} {\bibfield  {journal} {\bibinfo  {journal} {Phys.
  Rev. Lett.}\ }\textbf {\bibinfo {volume} {115}},\ \bibinfo {pages} {243602}
  (\bibinfo {year} {2015})}\BibitemShut {NoStop}%
\bibitem [{\citenamefont {Facchinetti}\ \emph {et~al.}(2016)\citenamefont
  {Facchinetti}, \citenamefont {Jenkins},\ and\ \citenamefont
  {Ruostekoski}}]{Facchinetti:2016}%
  \BibitemOpen
  \bibfield  {author} {\bibinfo {author} {\bibfnamefont {G.}~\bibnamefont
  {Facchinetti}}, \bibinfo {author} {\bibfnamefont {S.~D.}\ \bibnamefont
  {Jenkins}}, \ and\ \bibinfo {author} {\bibfnamefont {J.}~\bibnamefont
  {Ruostekoski}},\ }\bibfield  {title} {\enquote {\bibinfo {title} {Storing
  light with subradiant correlations in arrays of atoms},}\ }\href {\doibase
  10.1103/PhysRevLett.117.243601} {\bibfield  {journal} {\bibinfo  {journal}
  {Phys. Rev. Lett.}\ }\textbf {\bibinfo {volume} {117}},\ \bibinfo {pages}
  {243601} (\bibinfo {year} {2016})}\BibitemShut {NoStop}%
\bibitem [{\citenamefont {Jen}\ \emph {et~al.}(2016)\citenamefont {Jen},
  \citenamefont {Chang},\ and\ \citenamefont {Chen}}]{Jen:2016}%
  \BibitemOpen
  \bibfield  {author} {\bibinfo {author} {\bibfnamefont {H.~H.}\ \bibnamefont
  {Jen}}, \bibinfo {author} {\bibfnamefont {M.-S.}\ \bibnamefont {Chang}}, \
  and\ \bibinfo {author} {\bibfnamefont {Y.-C.}\ \bibnamefont {Chen}},\
  }\bibfield  {title} {\enquote {\bibinfo {title} {Cooperative single-photon
  subradiant states},}\ }\href {\doibase 10.1103/PhysRevA.94.013803} {\bibfield
   {journal} {\bibinfo  {journal} {Phys. Rev. A}\ }\textbf {\bibinfo {volume}
  {94}},\ \bibinfo {pages} {013803} (\bibinfo {year} {2016})}\BibitemShut
  {NoStop}%
\bibitem [{\citenamefont {Asenjo-Garcia}\ \emph {et~al.}(2017)\citenamefont
  {Asenjo-Garcia}, \citenamefont {Moreno-Cardoer}, \citenamefont {Albrecht},
  \citenamefont {Kimble},\ and\ \citenamefont {Chang}}]{Asenjo:2017}%
  \BibitemOpen
  \bibfield  {author} {\bibinfo {author} {\bibfnamefont {A.}~\bibnamefont
  {Asenjo-Garcia}}, \bibinfo {author} {\bibfnamefont {M.}~\bibnamefont
  {Moreno-Cardoer}}, \bibinfo {author} {\bibfnamefont {A.}~\bibnamefont
  {Albrecht}}, \bibinfo {author} {\bibfnamefont {H.~J.}\ \bibnamefont
  {Kimble}}, \ and\ \bibinfo {author} {\bibfnamefont {D.~E.}\ \bibnamefont
  {Chang}},\ }\bibfield  {title} {\enquote {\bibinfo {title} {Exponential
  improvement in photon storage fidelities using subradiance and ``selective
  radiance'' in atomic arrays},}\ }\href@noop {} {\bibfield  {journal}
  {\bibinfo  {journal} {Phys. Rev. X}\ }\textbf {\bibinfo {volume} {7}},\
  \bibinfo {pages} {031024} (\bibinfo {year} {2017})}\BibitemShut {NoStop}%
\bibitem [{\citenamefont {Holzinger}\ \emph {et~al.}(2020)\citenamefont
  {Holzinger}, \citenamefont {Plankensteiner}, \citenamefont {Ostermann},\ and\
  \citenamefont {Ritsch}}]{Holzinger:2020}%
  \BibitemOpen
  \bibfield  {author} {\bibinfo {author} {\bibfnamefont {R.}~\bibnamefont
  {Holzinger}}, \bibinfo {author} {\bibfnamefont {D.}~\bibnamefont
  {Plankensteiner}}, \bibinfo {author} {\bibfnamefont {L.}~\bibnamefont
  {Ostermann}}, \ and\ \bibinfo {author} {\bibfnamefont {H.}~\bibnamefont
  {Ritsch}},\ }\bibfield  {title} {\enquote {\bibinfo {title} {Nanoscale
  coherent light source},}\ }\href@noop {} {\bibfield  {journal} {\bibinfo
  {journal} {Phys. Rev. Lett.}\ }\textbf {\bibinfo {volume} {124}},\ \bibinfo
  {pages} {253603} (\bibinfo {year} {2020})}\BibitemShut {NoStop}%
\bibitem [{\citenamefont {Stephen}(1964)}]{Stephen:1964}%
  \BibitemOpen
  \bibfield  {author} {\bibinfo {author} {\bibfnamefont {M.~J.}\ \bibnamefont
  {Stephen}},\ }\bibfield  {title} {\enquote {\bibinfo {title} {First order
  dispersion forces},}\ }\href@noop {} {\bibfield  {journal} {\bibinfo
  {journal} {J. Chem. Phys.}\ }\textbf {\bibinfo {volume} {40}},\ \bibinfo
  {pages} {669} (\bibinfo {year} {1964})}\BibitemShut {NoStop}%
\bibitem [{\citenamefont {{DeVoe}}\ and\ \citenamefont
  {Brewer}(1996)}]{DeVoe:1996}%
  \BibitemOpen
  \bibfield  {author} {\bibinfo {author} {\bibfnamefont {R.~G.}\ \bibnamefont
  {{DeVoe}}}\ and\ \bibinfo {author} {\bibfnamefont {R.~G.}\ \bibnamefont
  {Brewer}},\ }\bibfield  {title} {\enquote {\bibinfo {title} {Observation of
  superradiant and subradiant spontaneous emission of two trapped ions},}\
  }\href@noop {} {\bibfield  {journal} {\bibinfo  {journal} {Phys. Rev. Lett.}\
  }\textbf {\bibinfo {volume} {76}},\ \bibinfo {pages} {2049--2052} (\bibinfo
  {year} {1996})}\BibitemShut {NoStop}%
\bibitem [{\citenamefont {Arecchi}\ and\ \citenamefont
  {Courtens}(1970)}]{Arecchi:1970}%
  \BibitemOpen
  \bibfield  {author} {\bibinfo {author} {\bibfnamefont {F.~T.}\ \bibnamefont
  {Arecchi}}\ and\ \bibinfo {author} {\bibfnamefont {E.}~\bibnamefont
  {Courtens}},\ }\bibfield  {title} {\enquote {\bibinfo {title} {Cooperative
  phenomena in resonant electromagnetic propagation},}\ }\href@noop {}
  {\bibfield  {journal} {\bibinfo  {journal} {Phys. Rev. A}\ }\textbf {\bibinfo
  {volume} {2}},\ \bibinfo {pages} {1730--1737} (\bibinfo {year}
  {1970})}\BibitemShut {NoStop}%
\bibitem [{\citenamefont {Rehler}\ and\ \citenamefont
  {Eberly}(1971)}]{Rehler:1971}%
  \BibitemOpen
  \bibfield  {author} {\bibinfo {author} {\bibfnamefont {N.~E.}\ \bibnamefont
  {Rehler}}\ and\ \bibinfo {author} {\bibfnamefont {J.~H.}\ \bibnamefont
  {Eberly}},\ }\bibfield  {title} {\enquote {\bibinfo {title}
  {Superradiance},}\ }\href@noop {} {\bibfield  {journal} {\bibinfo  {journal}
  {Phys. Rev. A}\ }\textbf {\bibinfo {volume} {3}},\ \bibinfo {pages}
  {1735--1751} (\bibinfo {year} {1971})}\BibitemShut {NoStop}%
\bibitem [{\citenamefont {{MacGillivray}}\ and\ \citenamefont
  {Feld}(1976)}]{MacGillivray:1976}%
  \BibitemOpen
  \bibfield  {author} {\bibinfo {author} {\bibfnamefont {J.~C.}\ \bibnamefont
  {{MacGillivray}}}\ and\ \bibinfo {author} {\bibfnamefont {M.~S.}\
  \bibnamefont {Feld}},\ }\bibfield  {title} {\enquote {\bibinfo {title}
  {Theory of superradiance in an extended, optically thick medium},}\
  }\href@noop {} {\bibfield  {journal} {\bibinfo  {journal} {Phys. Rev. A}\
  }\textbf {\bibinfo {volume} {14}},\ \bibinfo {pages} {1169--1189} (\bibinfo
  {year} {1976})}\BibitemShut {NoStop}%
\bibitem [{\citenamefont {Svidzinsky}\ \emph {et~al.}(2008)\citenamefont
  {Svidzinsky}, \citenamefont {Chang},\ and\ \citenamefont
  {Scully}}]{Svidzinsky:2008}%
  \BibitemOpen
  \bibfield  {author} {\bibinfo {author} {\bibfnamefont {A.~A.}\ \bibnamefont
  {Svidzinsky}}, \bibinfo {author} {\bibfnamefont {J.-T.}\ \bibnamefont
  {Chang}}, \ and\ \bibinfo {author} {\bibfnamefont {M.~O.}\ \bibnamefont
  {Scully}},\ }\bibfield  {title} {\enquote {\bibinfo {title} {Dynamical
  evolution of correlated spontaneous emission of a single photon from a
  uniformly excited cloud of {$N$} atoms},}\ }\href@noop {} {\bibfield
  {journal} {\bibinfo  {journal} {Phys. Rev. Lett.}\ }\textbf {\bibinfo
  {volume} {100}},\ \bibinfo {pages} {160504} (\bibinfo {year}
  {2008})}\BibitemShut {NoStop}%
\bibitem [{\citenamefont {Scully}\ and\ \citenamefont
  {Svidzinsky}(2009)}]{Scully:2009}%
  \BibitemOpen
  \bibfield  {author} {\bibinfo {author} {\bibfnamefont {M.~O.}\ \bibnamefont
  {Scully}}\ and\ \bibinfo {author} {\bibfnamefont {A.~A.}\ \bibnamefont
  {Svidzinsky}},\ }\bibfield  {title} {\enquote {\bibinfo {title} {The super of
  superradiance},}\ }\href@noop {} {\bibfield  {journal} {\bibinfo  {journal}
  {Science}\ }\textbf {\bibinfo {volume} {325}},\ \bibinfo {pages} {1510--1511}
  (\bibinfo {year} {2009})}\BibitemShut {NoStop}%
\bibitem [{\citenamefont {Ara\'{u}jo}\ \emph {et~al.}(2016)\citenamefont
  {Ara\'{u}jo}, \citenamefont {Kre\v{s}i\'{c}}, \citenamefont {Kaiser},\ and\
  \citenamefont {Guerin}}]{Araujo:2016}%
  \BibitemOpen
  \bibfield  {author} {\bibinfo {author} {\bibfnamefont {M.~O.}\ \bibnamefont
  {Ara\'{u}jo}}, \bibinfo {author} {\bibfnamefont {I.}~\bibnamefont
  {Kre\v{s}i\'{c}}}, \bibinfo {author} {\bibfnamefont {R.}~\bibnamefont
  {Kaiser}}, \ and\ \bibinfo {author} {\bibfnamefont {W.}~\bibnamefont
  {Guerin}},\ }\bibfield  {title} {\enquote {\bibinfo {title} {Superradiance in
  a large cloud of cold atoms in the linear-optics regime},}\ }\href@noop {}
  {\bibfield  {journal} {\bibinfo  {journal} {Phys. Rev. Lett.}\ }\textbf
  {\bibinfo {volume} {117}},\ \bibinfo {pages} {073002} (\bibinfo {year}
  {2016})}\BibitemShut {NoStop}%
\bibitem [{\citenamefont {Roof}\ \emph {et~al.}(2016)\citenamefont {Roof},
  \citenamefont {Kemp}, \citenamefont {Havey},\ and\ \citenamefont
  {Sokolov}}]{Roof:2016}%
  \BibitemOpen
  \bibfield  {author} {\bibinfo {author} {\bibfnamefont {S.~J.}\ \bibnamefont
  {Roof}}, \bibinfo {author} {\bibfnamefont {K.~J.}\ \bibnamefont {Kemp}},
  \bibinfo {author} {\bibfnamefont {M.~D.}\ \bibnamefont {Havey}}, \ and\
  \bibinfo {author} {\bibfnamefont {I.~M.}\ \bibnamefont {Sokolov}},\
  }\bibfield  {title} {\enquote {\bibinfo {title} {Observation of single-photon
  superradiance and the cooperative {Lamb} shift in an extended sample of cold
  atoms},}\ }\href@noop {} {\bibfield  {journal} {\bibinfo  {journal} {Phys.
  Rev. Lett.}\ }\textbf {\bibinfo {volume} {117}},\ \bibinfo {pages} {073003}
  (\bibinfo {year} {2016})}\BibitemShut {NoStop}%
\bibitem [{\citenamefont {Ortiz-Guti\'errez}\ \emph {et~al.}(2018)\citenamefont
  {Ortiz-Guti\'errez}, \citenamefont {Mu\~noz Mart\'{\i}nez}, \citenamefont
  {Barros}, \citenamefont {Morales}, \citenamefont {Moreira}, \citenamefont
  {Alves}, \citenamefont {Tieco}, \citenamefont {Saldanha},\ and\ \citenamefont
  {Felinto}}]{Ortiz:2018}%
  \BibitemOpen
  \bibfield  {author} {\bibinfo {author} {\bibfnamefont {L.}~\bibnamefont
  {Ortiz-Guti\'errez}}, \bibinfo {author} {\bibfnamefont {L.~F.}\ \bibnamefont
  {Mu\~noz Mart\'{\i}nez}}, \bibinfo {author} {\bibfnamefont {D.~F.}\
  \bibnamefont {Barros}}, \bibinfo {author} {\bibfnamefont {J.~E.~O.}\
  \bibnamefont {Morales}}, \bibinfo {author} {\bibfnamefont {R.~S.~N.}\
  \bibnamefont {Moreira}}, \bibinfo {author} {\bibfnamefont {N.~D.}\
  \bibnamefont {Alves}}, \bibinfo {author} {\bibfnamefont {A.~F.~G.}\
  \bibnamefont {Tieco}}, \bibinfo {author} {\bibfnamefont {P.~L.}\ \bibnamefont
  {Saldanha}}, \ and\ \bibinfo {author} {\bibfnamefont {D.}~\bibnamefont
  {Felinto}},\ }\bibfield  {title} {\enquote {\bibinfo {title} {Experimental
  fock-state superradiance},}\ }\href {\doibase 10.1103/PhysRevLett.120.083603}
  {\bibfield  {journal} {\bibinfo  {journal} {Phys. Rev. Lett.}\ }\textbf
  {\bibinfo {volume} {120}},\ \bibinfo {pages} {083603} (\bibinfo {year}
  {2018})}\BibitemShut {NoStop}%
\bibitem [{\citenamefont {Svidzinsky}\ and\ \citenamefont
  {Chang}(2008)}]{Svidzinsky:2008b}%
  \BibitemOpen
  \bibfield  {author} {\bibinfo {author} {\bibfnamefont {A.~A.}\ \bibnamefont
  {Svidzinsky}}\ and\ \bibinfo {author} {\bibfnamefont {{J.-T.}}\ \bibnamefont
  {Chang}},\ }\bibfield  {title} {\enquote {\bibinfo {title} {Cooperative
  spontaneous emission as a many-body eigenvalue problem},}\ }\href@noop {}
  {\bibfield  {journal} {\bibinfo  {journal} {Phys. Rev. A}\ }\textbf {\bibinfo
  {volume} {77}},\ \bibinfo {pages} {043833} (\bibinfo {year}
  {2008})}\BibitemShut {NoStop}%
\bibitem [{\citenamefont {Svidzinsky}\ \emph {et~al.}(2010)\citenamefont
  {Svidzinsky}, \citenamefont {Chang},\ and\ \citenamefont
  {Scully}}]{Svidzinsky:2010}%
  \BibitemOpen
  \bibfield  {author} {\bibinfo {author} {\bibfnamefont {A.~A.}\ \bibnamefont
  {Svidzinsky}}, \bibinfo {author} {\bibfnamefont {{J.-T.}}\ \bibnamefont
  {Chang}}, \ and\ \bibinfo {author} {\bibfnamefont {M.~O.}\ \bibnamefont
  {Scully}},\ }\bibfield  {title} {\enquote {\bibinfo {title} {Cooperative
  spontaneous emission of {$N$} atoms: Many-body eigenstates, the effect of
  virtual {Lamb} shift processes, and analogy with radiation of {$N$} classical
  oscillators},}\ }\href@noop {} {\bibfield  {journal} {\bibinfo  {journal}
  {Phys. Rev. A}\ }\textbf {\bibinfo {volume} {81}},\ \bibinfo {pages} {053821}
  (\bibinfo {year} {2010})}\BibitemShut {NoStop}%
\bibitem [{\citenamefont {Weiss}\ \emph {et~al.}(2021)\citenamefont {Weiss},
  \citenamefont {Cipris}, \citenamefont {Kaiser}, \citenamefont {Sokolov},\
  and\ \citenamefont {Guerin}}]{Weiss:2021}%
  \BibitemOpen
  \bibfield  {author} {\bibinfo {author} {\bibfnamefont {P.}~\bibnamefont
  {Weiss}}, \bibinfo {author} {\bibfnamefont {A.}~\bibnamefont {Cipris}},
  \bibinfo {author} {\bibfnamefont {R.}~\bibnamefont {Kaiser}}, \bibinfo
  {author} {\bibfnamefont {I.~M.}\ \bibnamefont {Sokolov}}, \ and\ \bibinfo
  {author} {\bibfnamefont {W.}~\bibnamefont {Guerin}},\ }\bibfield  {title}
  {\enquote {\bibinfo {title} {Superradiance as single scattering embedded in
  an effective medium},}\ }\href {\doibase 10.1103/PhysRevA.103.023702}
  {\bibfield  {journal} {\bibinfo  {journal} {Phys. Rev. A}\ }\textbf {\bibinfo
  {volume} {103}},\ \bibinfo {pages} {023702} (\bibinfo {year}
  {2021})}\BibitemShut {NoStop}%
\bibitem [{\citenamefont {Schilder}\ \emph {et~al.}(2016)\citenamefont
  {Schilder}, \citenamefont {Sauvan}, \citenamefont {Hugonin}, \citenamefont
  {Jennewein}, \citenamefont {Sortais}, \citenamefont {Browaeys},\ and\
  \citenamefont {Greffet}}]{Schilder:2016}%
  \BibitemOpen
  \bibfield  {author} {\bibinfo {author} {\bibfnamefont {N.~J.}\ \bibnamefont
  {Schilder}}, \bibinfo {author} {\bibfnamefont {C.}~\bibnamefont {Sauvan}},
  \bibinfo {author} {\bibfnamefont {J.-P.}\ \bibnamefont {Hugonin}}, \bibinfo
  {author} {\bibfnamefont {S.}~\bibnamefont {Jennewein}}, \bibinfo {author}
  {\bibfnamefont {Y.~R.~P.}\ \bibnamefont {Sortais}}, \bibinfo {author}
  {\bibfnamefont {A.}~\bibnamefont {Browaeys}}, \ and\ \bibinfo {author}
  {\bibfnamefont {J.-J.}\ \bibnamefont {Greffet}},\ }\bibfield  {title}
  {\enquote {\bibinfo {title} {Role of polaritonic modes on light scattering
  from a dense cloud of atoms},}\ }\href@noop {} {\bibfield  {journal}
  {\bibinfo  {journal} {Phys. Rev. A}\ }\textbf {\bibinfo {volume} {93}},\
  \bibinfo {pages} {063835} (\bibinfo {year} {2016})}\BibitemShut {NoStop}%
\bibitem [{\citenamefont {Cottier}\ \emph {et~al.}(2018)\citenamefont
  {Cottier}, \citenamefont {Kaiser},\ and\ \citenamefont
  {Bachelard}}]{Cottier:2018}%
  \BibitemOpen
  \bibfield  {author} {\bibinfo {author} {\bibfnamefont {F.}~\bibnamefont
  {Cottier}}, \bibinfo {author} {\bibfnamefont {R.}~\bibnamefont {Kaiser}}, \
  and\ \bibinfo {author} {\bibfnamefont {R.}~\bibnamefont {Bachelard}},\
  }\bibfield  {title} {\enquote {\bibinfo {title} {Rode of disorder in super-
  and subradiance of cold atomic clouds},}\ }\href@noop {} {\bibfield
  {journal} {\bibinfo  {journal} {Phys. Rev. A}\ }\textbf {\bibinfo {volume}
  {98}},\ \bibinfo {pages} {013622} (\bibinfo {year} {2018})}\BibitemShut
  {NoStop}%
\bibitem [{\citenamefont {Labeyrie}\ \emph {et~al.}(2003)\citenamefont
  {Labeyrie}, \citenamefont {Vaujour}, \citenamefont {M\"uller}, \citenamefont
  {Delande}, \citenamefont {Miniatura}, \citenamefont {Wilkowski},\ and\
  \citenamefont {Kaiser}}]{Labeyrie:2003}%
  \BibitemOpen
  \bibfield  {author} {\bibinfo {author} {\bibfnamefont {G.}~\bibnamefont
  {Labeyrie}}, \bibinfo {author} {\bibfnamefont {E.}~\bibnamefont {Vaujour}},
  \bibinfo {author} {\bibfnamefont {C.~A.}\ \bibnamefont {M\"uller}}, \bibinfo
  {author} {\bibfnamefont {D.}~\bibnamefont {Delande}}, \bibinfo {author}
  {\bibfnamefont {C.}~\bibnamefont {Miniatura}}, \bibinfo {author}
  {\bibfnamefont {D.}~\bibnamefont {Wilkowski}}, \ and\ \bibinfo {author}
  {\bibfnamefont {R.}~\bibnamefont {Kaiser}},\ }\bibfield  {title} {\enquote
  {\bibinfo {title} {Slow diffusion of light in a cold atomic cloud},}\
  }\href@noop {} {\bibfield  {journal} {\bibinfo  {journal} {Phys. Rev. Lett.}\
  }\textbf {\bibinfo {volume} {91}},\ \bibinfo {pages} {223904} (\bibinfo
  {year} {2003})}\BibitemShut {NoStop}%
\bibitem [{\citenamefont {van Rossum}\ and\ \citenamefont
  {Nieuwenhuizen}(1999)}]{Rossum:1999}%
  \BibitemOpen
  \bibfield  {author} {\bibinfo {author} {\bibfnamefont {M.~C.~W.}\
  \bibnamefont {van Rossum}}\ and\ \bibinfo {author} {\bibfnamefont {Th.~M.}\
  \bibnamefont {Nieuwenhuizen}},\ }\bibfield  {title} {\enquote {\bibinfo
  {title} {Multiple scattering of classical waves: microscopy, mesoscopy, and
  diffusion},}\ }\href@noop {} {\bibfield  {journal} {\bibinfo  {journal} {Rev.
  Mod. Phys.}\ }\textbf {\bibinfo {volume} {71}},\ \bibinfo {pages} {313--371}
  (\bibinfo {year} {1999})}\BibitemShut {NoStop}%
\bibitem [{\citenamefont {Foldy}(1945)}]{Foldy:1945}%
  \BibitemOpen
  \bibfield  {author} {\bibinfo {author} {\bibfnamefont {L.~L.}\ \bibnamefont
  {Foldy}},\ }\bibfield  {title} {\enquote {\bibinfo {title} {The multiple
  scattering of waves. {I. General} theory of isotropic scattering by randomly
  distributed scatterers},}\ }\href {\doibase 10.1103/PhysRev.67.107}
  {\bibfield  {journal} {\bibinfo  {journal} {Phys. Rev.}\ }\textbf {\bibinfo
  {volume} {67}},\ \bibinfo {pages} {107--119} (\bibinfo {year}
  {1945})}\BibitemShut {NoStop}%
\bibitem [{\citenamefont {Lax}(1951)}]{Lax:1951}%
  \BibitemOpen
  \bibfield  {author} {\bibinfo {author} {\bibfnamefont {M.}~\bibnamefont
  {Lax}},\ }\bibfield  {title} {\enquote {\bibinfo {title} {Multiple scattering
  of waves},}\ }\href@noop {} {\bibfield  {journal} {\bibinfo  {journal} {Rev.
  Mod. Phys.}\ }\textbf {\bibinfo {volume} {23}},\ \bibinfo {pages} {287--310}
  (\bibinfo {year} {1951})}\BibitemShut {NoStop}%
\bibitem [{\citenamefont {Javanainen}\ \emph {et~al.}(1999)\citenamefont
  {Javanainen}, \citenamefont {Ruostekoski}, \citenamefont {Vestergaard},\ and\
  \citenamefont {Francis}}]{Javanainen:1999}%
  \BibitemOpen
  \bibfield  {author} {\bibinfo {author} {\bibfnamefont {J.}~\bibnamefont
  {Javanainen}}, \bibinfo {author} {\bibfnamefont {J.}~\bibnamefont
  {Ruostekoski}}, \bibinfo {author} {\bibfnamefont {B.}~\bibnamefont
  {Vestergaard}}, \ and\ \bibinfo {author} {\bibfnamefont {M.~R.}\ \bibnamefont
  {Francis}},\ }\bibfield  {title} {\enquote {\bibinfo {title} {One-dimensional
  modeling of light propagation in dense and degenerate samples},}\ }\href@noop
  {} {\bibfield  {journal} {\bibinfo  {journal} {Phys. Rev. A}\ }\textbf
  {\bibinfo {volume} {59}},\ \bibinfo {pages} {649 -- 666} (\bibinfo {year}
  {1999})}\BibitemShut {NoStop}%
\bibitem [{\citenamefont {Rusek}\ \emph {et~al.}(2000)\citenamefont {Rusek},
  \citenamefont {Mostowski},\ and\ \citenamefont {Or\l{}owski}}]{Rusek:2000}%
  \BibitemOpen
  \bibfield  {author} {\bibinfo {author} {\bibfnamefont {M.}~\bibnamefont
  {Rusek}}, \bibinfo {author} {\bibfnamefont {J.}~\bibnamefont {Mostowski}}, \
  and\ \bibinfo {author} {\bibfnamefont {A.}~\bibnamefont {Or\l{}owski}},\
  }\bibfield  {title} {\enquote {\bibinfo {title} {Random green matrices: From
  proximity resonances to {Anderson} localization},}\ }\href@noop {} {\bibfield
   {journal} {\bibinfo  {journal} {Phys. Rev. A}\ }\textbf {\bibinfo {volume}
  {61}},\ \bibinfo {pages} {022704} (\bibinfo {year} {2000})}\BibitemShut
  {NoStop}%
\bibitem [{\citenamefont {Pinheiro}\ \emph {et~al.}(2004)\citenamefont
  {Pinheiro}, \citenamefont {Rusek}, \citenamefont {Orlowski},\ and\
  \citenamefont {van Tiggelen}}]{Pinheiro:2004}%
  \BibitemOpen
  \bibfield  {author} {\bibinfo {author} {\bibfnamefont {F.~A.}\ \bibnamefont
  {Pinheiro}}, \bibinfo {author} {\bibfnamefont {M.}~\bibnamefont {Rusek}},
  \bibinfo {author} {\bibfnamefont {A.}~\bibnamefont {Orlowski}}, \ and\
  \bibinfo {author} {\bibfnamefont {B.~A.}\ \bibnamefont {van Tiggelen}},\
  }\bibfield  {title} {\enquote {\bibinfo {title} {Probing {Anderson}
  localization of light via decay rate statistics},}\ }\href {\doibase
  10.1103/PhysRevE.69.026605} {\bibfield  {journal} {\bibinfo  {journal} {Phys.
  Rev. E}\ }\textbf {\bibinfo {volume} {69}},\ \bibinfo {pages} {026605}
  (\bibinfo {year} {2004})}\BibitemShut {NoStop}%
\bibitem [{\citenamefont {Fu}\ and\ \citenamefont {Berman}(2005)}]{Fu:2005}%
  \BibitemOpen
  \bibfield  {author} {\bibinfo {author} {\bibfnamefont {H.}~\bibnamefont
  {Fu}}\ and\ \bibinfo {author} {\bibfnamefont {P.~R.}\ \bibnamefont
  {Berman}},\ }\bibfield  {title} {\enquote {\bibinfo {title} {Microscopic
  theory of spontaneous decay in a dielectric},}\ }\href {\doibase
  10.1103/PhysRevA.72.022104} {\bibfield  {journal} {\bibinfo  {journal} {Phys.
  Rev. A}\ }\textbf {\bibinfo {volume} {72}},\ \bibinfo {pages} {022104}
  (\bibinfo {year} {2005})}\BibitemShut {NoStop}%
\bibitem [{\citenamefont {Courteille}\ \emph {et~al.}(2010)\citenamefont
  {Courteille}, \citenamefont {Bux}, \citenamefont {Lucioni}, \citenamefont
  {Lauber}, \citenamefont {Bienaim\'e}, \citenamefont {Kaiser},\ and\
  \citenamefont {Piovella}}]{Courteille:2010}%
  \BibitemOpen
  \bibfield  {author} {\bibinfo {author} {\bibfnamefont {Ph.~W.}\ \bibnamefont
  {Courteille}}, \bibinfo {author} {\bibfnamefont {S.}~\bibnamefont {Bux}},
  \bibinfo {author} {\bibfnamefont {E.}~\bibnamefont {Lucioni}}, \bibinfo
  {author} {\bibfnamefont {K.}~\bibnamefont {Lauber}}, \bibinfo {author}
  {\bibfnamefont {T.}~\bibnamefont {Bienaim\'e}}, \bibinfo {author}
  {\bibfnamefont {R.}~\bibnamefont {Kaiser}}, \ and\ \bibinfo {author}
  {\bibfnamefont {N.}~\bibnamefont {Piovella}},\ }\bibfield  {title} {\enquote
  {\bibinfo {title} {Modification of radiation pressure due to cooperative
  scattering of light},}\ }\href@noop {} {\bibfield  {journal} {\bibinfo
  {journal} {Eur. Phys. J. D.}\ }\textbf {\bibinfo {volume} {58}},\ \bibinfo
  {pages} {69--73} (\bibinfo {year} {2010})}\BibitemShut {NoStop}%
\bibitem [{\citenamefont {Kuznetsov}\ \emph {et~al.}(2011)\citenamefont
  {Kuznetsov}, \citenamefont {Roerich},\ and\ \citenamefont
  {Gladush}}]{Kuznetsov:2011}%
  \BibitemOpen
  \bibfield  {author} {\bibinfo {author} {\bibfnamefont {D.~V.}\ \bibnamefont
  {Kuznetsov}}, \bibinfo {author} {\bibfnamefont {Vl.~K.}\ \bibnamefont
  {Roerich}}, \ and\ \bibinfo {author} {\bibfnamefont {M.~G.}\ \bibnamefont
  {Gladush}},\ }\bibfield  {title} {\enquote {\bibinfo {title} {Local field and
  radiative relaxation rate in a dielectric medium},}\ }\href {\doibase
  10.1134/S1063776111100050} {\bibfield  {journal} {\bibinfo  {journal} {JETP}\
  }\textbf {\bibinfo {volume} {113}},\ \bibinfo {pages} {647--658} (\bibinfo
  {year} {2011})}\BibitemShut {NoStop}%
\bibitem [{\citenamefont {Skipetrov}\ and\ \citenamefont
  {Sokolov}(2014)}]{Skipetrov:2014}%
  \BibitemOpen
  \bibfield  {author} {\bibinfo {author} {\bibfnamefont {S.~E.}\ \bibnamefont
  {Skipetrov}}\ and\ \bibinfo {author} {\bibfnamefont {I.~M.}\ \bibnamefont
  {Sokolov}},\ }\bibfield  {title} {\enquote {\bibinfo {title} {Absence of
  {Anderson} localization of light in a random ensemble of point scatterers},}\
  }\href@noop {} {\bibfield  {journal} {\bibinfo  {journal} {Phys. Rev. Lett.}\
  }\textbf {\bibinfo {volume} {112}},\ \bibinfo {pages} {023905} (\bibinfo
  {year} {2014})}\BibitemShut {NoStop}%
\bibitem [{\citenamefont {Bellando}\ \emph {et~al.}(2014)\citenamefont
  {Bellando}, \citenamefont {Gero}, \citenamefont {Akkermans},\ and\
  \citenamefont {Kaiser}}]{Bellando:2014}%
  \BibitemOpen
  \bibfield  {author} {\bibinfo {author} {\bibfnamefont {L.}~\bibnamefont
  {Bellando}}, \bibinfo {author} {\bibfnamefont {A.}~\bibnamefont {Gero}},
  \bibinfo {author} {\bibfnamefont {E.}~\bibnamefont {Akkermans}}, \ and\
  \bibinfo {author} {\bibfnamefont {R.}~\bibnamefont {Kaiser}},\ }\bibfield
  {title} {\enquote {\bibinfo {title} {Cooperative effects and disorder: A
  scaling analysis of the spectrum of the effective atomic {Hamiltonian}},}\
  }\href@noop {} {\bibfield  {journal} {\bibinfo  {journal} {Phys. Rev. A}\
  }\textbf {\bibinfo {volume} {90}},\ \bibinfo {pages} {063822} (\bibinfo
  {year} {2014})}\BibitemShut {NoStop}%
\bibitem [{\citenamefont {Skipetrov}\ and\ \citenamefont
  {Sokolov}(2015)}]{Skipetrov:2015}%
  \BibitemOpen
  \bibfield  {author} {\bibinfo {author} {\bibfnamefont {S.~E.}\ \bibnamefont
  {Skipetrov}}\ and\ \bibinfo {author} {\bibfnamefont {I.~M.}\ \bibnamefont
  {Sokolov}},\ }\bibfield  {title} {\enquote {\bibinfo {title}
  {Magnetic-field-driven localization of light in a cold-atom gas},}\
  }\href@noop {} {\bibfield  {journal} {\bibinfo  {journal} {Phys. Rev. Lett.}\
  }\textbf {\bibinfo {volume} {114}},\ \bibinfo {pages} {053902} (\bibinfo
  {year} {2015})}\BibitemShut {NoStop}%
\bibitem [{\citenamefont {Kuraptsev}\ and\ \citenamefont
  {Sokolov}(2015)}]{Kuraptsev:2015}%
  \BibitemOpen
  \bibfield  {author} {\bibinfo {author} {\bibfnamefont {A.~S.}\ \bibnamefont
  {Kuraptsev}}\ and\ \bibinfo {author} {\bibfnamefont {I.~M.}\ \bibnamefont
  {Sokolov}},\ }\bibfield  {title} {\enquote {\bibinfo {title} {Reflection of
  resonant light from a plane surface of an ensemble of motionless point
  scatters: Quantum microscopic approach},}\ }\href {\doibase
  10.1103/PhysRevA.91.053822} {\bibfield  {journal} {\bibinfo  {journal} {Phys.
  Rev. A}\ }\textbf {\bibinfo {volume} {91}},\ \bibinfo {pages} {053822}
  (\bibinfo {year} {2015})}\BibitemShut {NoStop}%
\bibitem [{\citenamefont {Sokolov}\ \emph {et~al.}(2011)\citenamefont
  {Sokolov}, \citenamefont {Kupriyanov},\ and\ \citenamefont
  {Havey}}]{Sokolov:2011}%
  \BibitemOpen
  \bibfield  {author} {\bibinfo {author} {\bibfnamefont {I.M.}\ \bibnamefont
  {Sokolov}}, \bibinfo {author} {\bibfnamefont {D.V.}\ \bibnamefont
  {Kupriyanov}}, \ and\ \bibinfo {author} {\bibfnamefont {M.D.}\ \bibnamefont
  {Havey}},\ }\bibfield  {title} {\enquote {\bibinfo {title} {Microscopic
  theory of scattering of weak electromagnetic radiation by a dense ensemble of
  ultracold atoms},}\ }\href@noop {} {\bibfield  {journal} {\bibinfo  {journal}
  {JETP}\ }\textbf {\bibinfo {volume} {112}},\ \bibinfo {pages} {246} (\bibinfo
  {year} {2011})}\BibitemShut {NoStop}%
\bibitem [{\citenamefont {Kuraptsev}\ \emph {et~al.}(2017)\citenamefont
  {Kuraptsev}, \citenamefont {Sokolov},\ and\ \citenamefont
  {Havey}}]{Kuraptsev:2017}%
  \BibitemOpen
  \bibfield  {author} {\bibinfo {author} {\bibfnamefont {A.~S.}\ \bibnamefont
  {Kuraptsev}}, \bibinfo {author} {\bibfnamefont {I.}~\bibnamefont {Sokolov}},
  \ and\ \bibinfo {author} {\bibfnamefont {M.~D.}\ \bibnamefont {Havey}},\
  }\bibfield  {title} {\enquote {\bibinfo {title} {Angular distribution of
  single photon superradiance in a dilute and cold atomic ensemble},}\
  }\href@noop {} {\bibfield  {journal} {\bibinfo  {journal} {Phys. Rev. A}\
  }\textbf {\bibinfo {volume} {96}},\ \bibinfo {pages} {023830} (\bibinfo
  {year} {2017})}\BibitemShut {NoStop}%
\bibitem [{\citenamefont {Ara\'ujo}\ \emph {et~al.}(2018)\citenamefont
  {Ara\'ujo}, \citenamefont {Guerin},\ and\ \citenamefont
  {Kaiser}}]{Araujo:2018}%
  \BibitemOpen
  \bibfield  {author} {\bibinfo {author} {\bibfnamefont {M.~O.}\ \bibnamefont
  {Ara\'ujo}}, \bibinfo {author} {\bibfnamefont {W.}~\bibnamefont {Guerin}}, \
  and\ \bibinfo {author} {\bibfnamefont {R.}~\bibnamefont {Kaiser}},\
  }\bibfield  {title} {\enquote {\bibinfo {title} {Decay dynamics in the
  coupled-dipole model},}\ }\href@noop {} {\bibfield  {journal} {\bibinfo
  {journal} {J. Mod. Opt.}\ }\textbf {\bibinfo {volume} {65}},\ \bibinfo
  {pages} {1345--1354} (\bibinfo {year} {2018})}\BibitemShut {NoStop}%
\bibitem [{\citenamefont {Gero}\ and\ \citenamefont
  {Akkermans}(2006)}]{Gero:2006}%
  \BibitemOpen
  \bibfield  {author} {\bibinfo {author} {\bibfnamefont {A.}~\bibnamefont
  {Gero}}\ and\ \bibinfo {author} {\bibfnamefont {E.}~\bibnamefont
  {Akkermans}},\ }\bibfield  {title} {\enquote {\bibinfo {title} {Effect of
  superradiance on transport of diffusing photons in cold atomic gases},}\
  }\href@noop {} {\bibfield  {journal} {\bibinfo  {journal} {Phys. Rev. Lett.}\
  }\textbf {\bibinfo {volume} {96}},\ \bibinfo {pages} {093601} (\bibinfo
  {year} {2006})}\BibitemShut {NoStop}%
\bibitem [{\citenamefont {Weiss}\ \emph {et~al.}(2018)\citenamefont {Weiss},
  \citenamefont {Ara\'ujo}, \citenamefont {Kaiser},\ and\ \citenamefont
  {Guerin}}]{Weiss:2018}%
  \BibitemOpen
  \bibfield  {author} {\bibinfo {author} {\bibfnamefont {P.}~\bibnamefont
  {Weiss}}, \bibinfo {author} {\bibfnamefont {M.~O.}\ \bibnamefont {Ara\'ujo}},
  \bibinfo {author} {\bibfnamefont {R.}~\bibnamefont {Kaiser}}, \ and\ \bibinfo
  {author} {\bibfnamefont {W.}~\bibnamefont {Guerin}},\ }\bibfield  {title}
  {\enquote {\bibinfo {title} {Subradiance and radiation trapping in cold
  atoms},}\ }\href {\doibase doi.org/10.1088/1367-2630/aac5d0} {\bibfield
  {journal} {\bibinfo  {journal} {New. J. Phys.}\ }\textbf {\bibinfo {volume}
  {20}},\ \bibinfo {pages} {063024} (\bibinfo {year} {2018})}\BibitemShut
  {NoStop}%
\bibitem [{\citenamefont {Cao}(2003)}]{Cao:2003}%
  \BibitemOpen
  \bibfield  {author} {\bibinfo {author} {\bibfnamefont {H.}~\bibnamefont
  {Cao}},\ }\bibfield  {title} {\enquote {\bibinfo {title} {Lasing in random
  media},}\ }\href@noop {} {\bibfield  {journal} {\bibinfo  {journal} {Waves
  Random Media}\ }\textbf {\bibinfo {volume} {13}},\ \bibinfo {pages} {R1--R39}
  (\bibinfo {year} {2003})}\BibitemShut {NoStop}%
\bibitem [{\citenamefont {Guerin}\ and\ \citenamefont
  {Kaiser}(2017)}]{Guerin:2017b}%
  \BibitemOpen
  \bibfield  {author} {\bibinfo {author} {\bibfnamefont {W.}~\bibnamefont
  {Guerin}}\ and\ \bibinfo {author} {\bibfnamefont {R.}~\bibnamefont
  {Kaiser}},\ }\bibfield  {title} {\enquote {\bibinfo {title} {Population of
  collective modes in light scattering by many atoms},}\ }\href@noop {}
  {\bibfield  {journal} {\bibinfo  {journal} {Phys. Rev. A}\ }\textbf {\bibinfo
  {volume} {95}},\ \bibinfo {pages} {053865} (\bibinfo {year}
  {2017})}\BibitemShut {NoStop}%
\bibitem [{\citenamefont {Bozhokin}\ and\ \citenamefont
  {Sokolov}(2018)}]{Bozhokin:2018}%
  \BibitemOpen
  \bibfield  {author} {\bibinfo {author} {\bibfnamefont {S.~V.}\ \bibnamefont
  {Bozhokin}}\ and\ \bibinfo {author} {\bibfnamefont {I.~M.}\ \bibnamefont
  {Sokolov}},\ }\bibfield  {title} {\enquote {\bibinfo {title} {Comparison of
  the wavelet and {Gabor} transforms in the spectral analysis of nonstationary
  signals},}\ }\href@noop {} {\bibfield  {journal} {\bibinfo  {journal} {Tech.
  Phys.}\ }\textbf {\bibinfo {volume} {63}},\ \bibinfo {pages} {1711--1717}
  (\bibinfo {year} {2018})}\BibitemShut {NoStop}%
\bibitem [{\citenamefont {Cipris}\ \emph {et~al.}()\citenamefont {Cipris},
  \citenamefont {Bachelard}, \citenamefont {Kaiser},\ and\ \citenamefont
  {Guerin}}]{Cipris:2021b}%
  \BibitemOpen
  \bibfield  {author} {\bibinfo {author} {\bibfnamefont {A.}~\bibnamefont
  {Cipris}}, \bibinfo {author} {\bibfnamefont {R.}~\bibnamefont {Bachelard}},
  \bibinfo {author} {\bibfnamefont {R.}~\bibnamefont {Kaiser}}, \ and\ \bibinfo
  {author} {\bibfnamefont {W.}~\bibnamefont {Guerin}},\ }\bibfield  {title}
  {\enquote {\bibinfo {title} {Van der {Waals} dephasing for {Dicke}
  subradiance in cold atomic clouds},}\ } \href {\doibase
  10.1103/PhysRevA.103.033714} {\bibfield  {journal} {\bibinfo  {journal} {Phys.
  Rev. A}\ }\textbf {\bibinfo {volume} {103}},\ \bibinfo {pages} {033714}
  (\bibinfo {year} {2021})} \BibitemShut
  {NoStop}%
\bibitem [{\citenamefont {Caires}\ \emph {et~al.}(2004)\citenamefont {Caires},
  \citenamefont {Telles}, \citenamefont {Mancini}, \citenamefont {Marcassa},
  \citenamefont {Bagnato}, \citenamefont {Wilkowski},\ and\ \citenamefont
  {Kaiser}}]{Caires:2004}%
  \BibitemOpen
  \bibfield  {author} {\bibinfo {author} {\bibfnamefont {A.~R.~L.}\
  \bibnamefont {Caires}}, \bibinfo {author} {\bibfnamefont {G.~D.}\
  \bibnamefont {Telles}}, \bibinfo {author} {\bibfnamefont {M.~W.}\
  \bibnamefont {Mancini}}, \bibinfo {author} {\bibfnamefont {L.~G.}\
  \bibnamefont {Marcassa}}, \bibinfo {author} {\bibfnamefont {V.~S.}\
  \bibnamefont {Bagnato}}, \bibinfo {author} {\bibfnamefont {D.}~\bibnamefont
  {Wilkowski}}, \ and\ \bibinfo {author} {\bibfnamefont {R.}~\bibnamefont
  {Kaiser}},\ }\bibfield  {title} {\enquote {\bibinfo {title} {Intensity
  dependence for trap loss rate in a magneto-optical trap of strontium},}\
  }\href@noop {} {\bibfield  {journal} {\bibinfo  {journal} {Braz. J. Phys.}\
  }\textbf {\bibinfo {volume} {34}},\ \bibinfo {pages} {1504} (\bibinfo {year}
  {2004})}\BibitemShut {NoStop}%
\bibitem [{\citenamefont {Fuhrmanek}\ \emph {et~al.}(2012)\citenamefont
  {Fuhrmanek}, \citenamefont {Bourgain}, \citenamefont {Sortais},\ and\
  \citenamefont {Browaeys}}]{Fuhrmanek:2012}%
  \BibitemOpen
  \bibfield  {author} {\bibinfo {author} {\bibfnamefont {A.}~\bibnamefont
  {Fuhrmanek}}, \bibinfo {author} {\bibfnamefont {R.}~\bibnamefont {Bourgain}},
  \bibinfo {author} {\bibfnamefont {Y.~R.~P.}\ \bibnamefont {Sortais}}, \ and\
  \bibinfo {author} {\bibfnamefont {A.}~\bibnamefont {Browaeys}},\ }\bibfield
  {title} {\enquote {\bibinfo {title} {Light-assisted collisions between a few
  cold atoms in a microscopic dipole trap},}\ }\href {\doibase
  10.1103/PhysRevA.85.062708} {\bibfield  {journal} {\bibinfo  {journal} {Phys.
  Rev. A}\ }\textbf {\bibinfo {volume} {85}},\ \bibinfo {pages} {062708}
  (\bibinfo {year} {2012})}\BibitemShut {NoStop}%
\bibitem [{\citenamefont {Weiss}\ \emph {et~al.}(2019)\citenamefont {Weiss},
  \citenamefont {Cipris}, \citenamefont {Ara\'ujo}, \citenamefont {Kaiser},\
  and\ \citenamefont {Guerin}}]{Weiss:2019}%
  \BibitemOpen
  \bibfield  {author} {\bibinfo {author} {\bibfnamefont {P.}~\bibnamefont
  {Weiss}}, \bibinfo {author} {\bibfnamefont {A.}~\bibnamefont {Cipris}},
  \bibinfo {author} {\bibfnamefont {M.~O.}\ \bibnamefont {Ara\'ujo}}, \bibinfo
  {author} {\bibfnamefont {R.}~\bibnamefont {Kaiser}}, \ and\ \bibinfo {author}
  {\bibfnamefont {W.}~\bibnamefont {Guerin}},\ }\bibfield  {title} {\enquote
  {\bibinfo {title} {Robustness of {Dicke} subradiance against thermal
  decoherence},}\ }\href@noop {} {\bibfield  {journal} {\bibinfo  {journal}
  {Phys. Rev. A}\ }\textbf {\bibinfo {volume} {100}},\ \bibinfo {pages}
  {033833} (\bibinfo {year} {2019})}\BibitemShut {NoStop}%
\bibitem [{\citenamefont {Kuraptsev}\ and\ \citenamefont
  {Sokolov}(2020)}]{Kuraptsev:2020}%
  \BibitemOpen
  \bibfield  {author} {\bibinfo {author} {\bibfnamefont {A.~S.}\ \bibnamefont
  {Kuraptsev}}\ and\ \bibinfo {author} {\bibfnamefont {I.~M.}\ \bibnamefont
  {Sokolov}},\ }\bibfield  {title} {\enquote {\bibinfo {title} {Influence of
  atomic motion on the collective effects in dense and cold atomic
  ensembles},}\ }\href {\doibase 10.1103/PhysRevA.101.033602} {\bibfield
  {journal} {\bibinfo  {journal} {Phys. Rev. A}\ }\textbf {\bibinfo {volume}
  {101}},\ \bibinfo {pages} {033602} (\bibinfo {year} {2020})}\BibitemShut
  {NoStop}%
\bibitem [{\citenamefont {Gisbert}\ \emph {et~al.}(2019)\citenamefont
  {Gisbert}, \citenamefont {Piovella},\ and\ \citenamefont
  {Bachelard}}]{Gisbert:2019}%
  \BibitemOpen
  \bibfield  {author} {\bibinfo {author} {\bibfnamefont {A.~T.}\ \bibnamefont
  {Gisbert}}, \bibinfo {author} {\bibfnamefont {N.}~\bibnamefont {Piovella}}, \
  and\ \bibinfo {author} {\bibfnamefont {R.}~\bibnamefont {Bachelard}},\
  }\bibfield  {title} {\enquote {\bibinfo {title} {Stochastic heating and
  self-induced cooling in optically bound pairs of atoms},}\ }\href {\doibase
  10.1103/PhysRevA.99.013619} {\bibfield  {journal} {\bibinfo  {journal} {Phys.
  Rev. A}\ }\textbf {\bibinfo {volume} {99}},\ \bibinfo {pages} {013619}
  (\bibinfo {year} {2019})}\BibitemShut {NoStop}%
\bibitem [{\citenamefont {Labeyrie}\ \emph {et~al.}(2005)\citenamefont
  {Labeyrie}, \citenamefont {Kaiser},\ and\ \citenamefont
  {Delande}}]{Labeyrie:2005}%
  \BibitemOpen
  \bibfield  {author} {\bibinfo {author} {\bibfnamefont {G.}~\bibnamefont
  {Labeyrie}}, \bibinfo {author} {\bibfnamefont {R.}~\bibnamefont {Kaiser}}, \
  and\ \bibinfo {author} {\bibfnamefont {D.}~\bibnamefont {Delande}},\
  }\bibfield  {title} {\enquote {\bibinfo {title} {Radiation trapping in a cold
  atomic gas},}\ }\href@noop {} {\bibfield  {journal} {\bibinfo  {journal}
  {Appl. Phys. B}\ }\textbf {\bibinfo {volume} {81}},\ \bibinfo {pages}
  {1001--1008} (\bibinfo {year} {2005})}\BibitemShut {NoStop}%
\bibitem [{\citenamefont {Pierrat}\ \emph {et~al.}(2009)\citenamefont
  {Pierrat}, \citenamefont {Gr\'emaud},\ and\ \citenamefont
  {Delande}}]{Pierrat:2009}%
  \BibitemOpen
  \bibfield  {author} {\bibinfo {author} {\bibfnamefont {R.}~\bibnamefont
  {Pierrat}}, \bibinfo {author} {\bibfnamefont {B.}~\bibnamefont {Gr\'emaud}},
  \ and\ \bibinfo {author} {\bibfnamefont {D.}~\bibnamefont {Delande}},\
  }\bibfield  {title} {\enquote {\bibinfo {title} {Enhancement of radiation
  trapping for quasiresonant scatterers at low temperature},}\ }\href@noop {}
  {\bibfield  {journal} {\bibinfo  {journal} {Phys. Rev. A}\ }\textbf {\bibinfo
  {volume} {80}},\ \bibinfo {pages} {013831} (\bibinfo {year}
  {2009})}\BibitemShut {NoStop}%
\bibitem [{\citenamefont {Datsyuk}\ and\ \citenamefont
  {Sokolov}(2006)}]{Datsyuk:2006}%
  \BibitemOpen
  \bibfield  {author} {\bibinfo {author} {\bibfnamefont {V.~M.}\ \bibnamefont
  {Datsyuk}}\ and\ \bibinfo {author} {\bibfnamefont {I.~M.}\ \bibnamefont
  {Sokolov}},\ }\bibfield  {title} {\enquote {\bibinfo {title} {Coherent
  backscattering under conditions of pulsed radiation trapping},}\ }\href@noop
  {} {\bibfield  {journal} {\bibinfo  {journal} {JETP}\ }\textbf {\bibinfo
  {volume} {102}},\ \bibinfo {pages} {724--736} (\bibinfo {year}
  {2006})}\BibitemShut {NoStop}%
\bibitem [{\citenamefont {Chab\'e}\ \emph {et~al.}(2014)\citenamefont
  {Chab\'e}, \citenamefont {Rouabah}, \citenamefont {Bellando}, \citenamefont
  {Bienaim\'e}, \citenamefont {Piovella}, \citenamefont {Bachelard},\ and\
  \citenamefont {Kaiser}}]{Chabe:2014}%
  \BibitemOpen
  \bibfield  {author} {\bibinfo {author} {\bibfnamefont {J.}~\bibnamefont
  {Chab\'e}}, \bibinfo {author} {\bibfnamefont {M.~T.}\ \bibnamefont
  {Rouabah}}, \bibinfo {author} {\bibfnamefont {L.}~\bibnamefont {Bellando}},
  \bibinfo {author} {\bibfnamefont {T.}~\bibnamefont {Bienaim\'e}}, \bibinfo
  {author} {\bibfnamefont {N.}~\bibnamefont {Piovella}}, \bibinfo {author}
  {\bibfnamefont {R.}~\bibnamefont {Bachelard}}, \ and\ \bibinfo {author}
  {\bibfnamefont {R.}~\bibnamefont {Kaiser}},\ }\bibfield  {title} {\enquote
  {\bibinfo {title} {Coherent and incoherent multiple scattering},}\
  }\href@noop {} {\bibfield  {journal} {\bibinfo  {journal} {Phys. Rev. A}\
  }\textbf {\bibinfo {volume} {89}},\ \bibinfo {pages} {043833} (\bibinfo
  {year} {2014})}\BibitemShut {NoStop}%
\bibitem [{\citenamefont {Sokolov}\ and\ \citenamefont
  {Guerin}(2019)}]{Sokolov:2019}%
  \BibitemOpen
  \bibfield  {author} {\bibinfo {author} {\bibfnamefont {I.~M.}\ \bibnamefont
  {Sokolov}}\ and\ \bibinfo {author} {\bibfnamefont {W.}~\bibnamefont
  {Guerin}},\ }\bibfield  {title} {\enquote {\bibinfo {title} {Comparison of
  three approaches to light scattering by dilute cold atomic ensembles},}\
  }\href@noop {} {\bibfield  {journal} {\bibinfo  {journal} {J. Opt. Soc. Am.
  B}\ }\textbf {\bibinfo {volume} {36}},\ \bibinfo {pages} {2030--3037}
  (\bibinfo {year} {2019})}\BibitemShut {NoStop}%
\bibitem [{\citenamefont {Fofanov}\ \emph {et~al.}(2020)\citenamefont
  {Fofanov}, \citenamefont {Manoylov}, \citenamefont {Zarutskiy},\ and\
  \citenamefont {Kuraptsev}}]{Fofanov:2020}%
  \BibitemOpen
  \bibfield  {author} {\bibinfo {author} {\bibfnamefont {Y.A.}\ \bibnamefont
  {Fofanov}}, \bibinfo {author} {\bibfnamefont {V.V.}\ \bibnamefont
  {Manoylov}}, \bibinfo {author} {\bibfnamefont {I.V.}\ \bibnamefont
  {Zarutskiy}}, \ and\ \bibinfo {author} {\bibfnamefont {A.S}\ \bibnamefont
  {Kuraptsev}},\ }\bibfield  {title} {\enquote {\bibinfo {title} {Laser
  polarization-optical diagnostics of ordered objects and structures},}\ }\href
  {\doibase 10.3103/S1062873820030089} {\bibfield  {journal} {\bibinfo
  {journal} {Bull. Russ. Acad. Sci. Phys.}\ }\textbf {\bibinfo {volume} {84}},\
  \bibinfo {pages} {263--266} (\bibinfo {year} {2020})}\BibitemShut {NoStop}%
\end{thebibliography}
%

\end{document}